\documentclass[pdflatex,sn-mathphys-num]{sn-jnl}

\usepackage{graphicx}%
\usepackage{multirow}%
\usepackage{amsmath,amssymb,amsfonts}%
\usepackage{amsthm}%
\usepackage{mathrsfs}%
\usepackage[title]{appendix}%
\usepackage{xcolor}%
\usepackage{soul}\sethlcolor{yellow}\soulregister{\cite}{7}\soulregister{\ref}{7}\soulregister{\textit}{7}
\usepackage{textcomp}%
\usepackage{manyfoot}%
\usepackage{booktabs}%
\usepackage{algorithm}%
\usepackage{algorithmicx}%
\usepackage{algpseudocode}%
\usepackage{listings}%
\usepackage{pifont}
\usepackage{subfig}
\usepackage{color, colortbl}
\usepackage{booktabs}
\definecolor{g1}{HTML}{D6D5C1} 
\definecolor{g2}{HTML}{FAFAF6}
\definecolor{g3}{HTML}{ECEBE2}

\newcommand*{\belowrulesepcolor}[1]{%
  \noalign{%
    \kern-\belowrulesep 
    \begingroup 
      \color{#1}%
      \hrule height\belowrulesep 
    \endgroup 
  }%
} 
\newcommand*{\aboverulesepcolor}[1]{%
  \noalign{%
    \begingroup 
      \color{#1}%
      \hrule height\aboverulesep 
    \endgroup 
    \kern-\aboverulesep 
  }%
} 

\theoremstyle{thmstyleone}%
\theoremstyle{thmstyletwo}%

\theoremstyle{thmstylethree}%

\begin{document}

\title[Cyst-X: A Multicenter MRI Benchmark for Pancreatic Cyst Risk Stratification and Cross-Institutional AI]{Cyst-X: {A Multicenter MRI Benchmark for Pancreatic Cyst Risk Stratification and Cross-Institutional AI}}


\author[1]{Hongyi Pan}\equalcont{These authors contributed equally to this work.}
\author[1]{Gorkem Durak}\equalcont{These authors contributed equally to this work.}
\author[1]{Halil Ertugrul Aktas}
\author[1]{Ziliang Hong}
\author[1]{Zhixiang Wang}
\author[1]{Elif Keles}
\affil[1]{Machine \& Hybrid Intelligence Lab, Department of Radiology, Northwestern University, Chicago, IL, USA}%
\author[2]{Deniz Seyithanoglu}
\affil[2]{Istanbul Faculty of Medicine, Istanbul University, Istanbul, Turkey}
\author[1]{Zheyuan Zhang}
\author[2]{Alpay Medetalibeyoglu}
\author[1]{Andrea Mia Bejar}
\author[2]{Yavuz Taktak}
\author[2]{Gulbiz Dagoglu Kartal}
\author[2]{Mehmet Sukru Erturk}
\author[2]{Timurhan Cebeci}
\affil[3]{Department of Biomedical Engineering and Radiology, University of Wisconsin-Madison, Madison, WI, USA}
\author[1]{Yury Velichko}
\author[4]{Lili Zhao}
\affil[4]{Department of Preventive Medicine, Northwestern University, Chicago, IL, USA}
\author[5]{Emil Agarunov}
\affil[5]{Division of Gastroenterology and Hepatology, New York University, New York, NY, USA}
\author[6]{Federica Proietto Salanitri}
\affil[6]{Department of Electrical, Electronic and Computer Engineering, University of Catania, Catania, Italy}
\author[6]{Concetto Spampinato}
\author[3]{Pallavi Tiwari}
\author[7]{Ziyue Xu}
\affil[7]{NVIDIA, Bethesda, MD, USA}
\author[8]{Sachin Jambawalikar}
\affil[8]{Department of Radiology, Columbia University, New York, NY, USA}
\author[9]{Ivo G. Schoots}
\affil[9]{Department of Radiology and Nuclear Medicine, Erasmus Medical Center, Rotterdam, Netherlands}
\author[10]{Marco J. Bruno}
\affil[10]{Department of Gastroenterology and Hepatology, Erasmus Medical Center, Rotterdam, Netherlands}
\author[11]{Chenchan Huang}
\affil[11]{Department of Radiology, New York University, New York, NY, USA}
\author[12]{Candice W. Bolan}
\affil[12]{Department of Radiology, Mayo Clinic Florida, Jacksonville, FL, USA}
\author[5]{Tamas Gonda}
\author[1]{Frank H. Miller}
\author[13]{Rajesh N. Keswani}
\affil[13]{Department of Gastroenterology and Hepatology, Northwestern University, Chicago, IL, USA}
\author[14]{Michael B. Wallace}
\affil[14]{Division of Gastroenterology and Hepatology, Mayo Clinic Florida, Jacksonville, FL, USA}
\author*[1]{Ulas Bagci}\email{ulas.bagci@northwestern.edu}

\abstract{{Intraductal papillary mucinous neoplasms (IPMNs) are pancreatic cancer precursors, and guideline-based risk stratification both over-treats indolent lesions and misses progression. We present \textit{Cyst-X}, a public benchmark of 1,461 T1-weighted (T1W) and T2-weighted (T2W) MRI volumes from 764 patients at seven centers, with expert pancreas masks, risk labels, splits and reference AI risk-prediction models. \textit{PanSegNet} segmented the pancreas with Dice of 86.34\% (T1W) and 89.26\% (T2W) on held-out centers. For the primary task, high-risk versus low-risk IPMN, a 3D DenseNet-121 achieved cross-validated AUCs of 74.59\% and 73.45\%; complementary broad binary and three-class tasks reached 78.67--81.28\% and 73.47--75.76\%. Because patient images often cannot be pooled across institutions, we also simulated federated learning across the seven centers. On T2W MRI, FedProx was similar to centralized training in center-averaged external AUC (71.00\% versus 71.18\%) but not in pooled AUC (62.35\% versus 77.49\%). Radiologist comparisons showed cohort-dependent sensitivity--specificity tradeoffs. \textit{Cyst-X} supports reproducible method development and identifies within-IPMN risk discrimination and cross-center score comparability as open challenges.}}




\maketitle

\section{Main}
Pancreatic ductal adenocarcinoma (PDAC) carries one of the worst prognoses in oncology, with most patients diagnosed at stages no longer amenable to curative resection~\cite{stoffel2023pancreatic}. By 2030, pancreatic cancer is projected to become the second leading cause of cancer death in 
{the United States~\cite{rahib2014projecting}}. Long-term survival remains possible when malignancy is detected before invasive transformation~\cite{sohn2004intraductal}
, which makes the small subset of identifiable pre-malignant lesions disproportionately important. Among pancreatic cystic lesions, intraductal papillary mucinous neoplasms (IPMNs) {are the largest pre-malignant subset}: roughly 15 to 20\% of pancreatic cancers arise from them~\cite{tanaka2017revisions}, and contemporary Magnetic Resonance Imaging (MRI) surveillance detects pancreatic cystic lesions in up to 49\% of older adults~\cite{ohtsuka2024international}. The clinical question is therefore not whether to detect IPMNs (modern cross-sectional imaging already does) but how to stratify, among the lesions detected, those whose malignant potential justifies surgical resection~\cite{corral2019deep,yao2023review}.

{The management of pancreatic cysts is determined by their risk of malignancy, which is classified as likely benign, low risk for malignancy, or potentially containing high-grade dysplasia or invasive carcinoma.}
Current consensus guidelines, including the Kyoto, Fukuoka, and American Gastroenterological Association (AGA) criteria, translate imaging and clinical features into binary surgical recommendations. {However, these recommendations agree imperfectly with final pathology, and the errors run in two clinically important directions}~\cite{heckler2017sendai,ohtsuka2024international,gonda2024pancreatic}. 
Meta-analysis of guideline-based management of branch-duct IPMN has identified limited specificity, highlighting the potential for surgical overtreatment~\cite{heckler2017sendai}.
{Conversely, initially low-risk IPMNs can develop worrisome features or high-risk stigmata during surveillance, and malignant transformation is a possibility~\cite{lattimore2023long}.}
The resulting clinical paradox, simultaneous over-treatment of indolent disease and under-detection of evolving malignancy, has motivated repeated calls to supplement guideline-based stratification with quantitative imaging analysis~\cite{robles2016accuracy,maggi2018pancreatic,bulcke2021evaluating,romutis2023burden}. 

MRI is the preferred non-invasive {imaging} modality for characterizing pancreatic cystic lesions. Compared with computed tomography (CT), MRI and magnetic resonance cholangiopancreatography offer superior sensitivity for duct communication~\cite{ohtsuka2024international}, mural nodules and multifocal disease, and they avoid the cumulative ionizing radiation exposure~\cite{seyithanoglu2024advances,megibow2017management,ohtsuka2024international} that is a concern in patients undergoing long-term surveillance~\cite{vege2015american,megibow2017management,seyithanoglu2024advances}. Yet artificial intelligence (AI) development for pancreatic imaging has unfolded largely on CT, most prominently in recent detection systems for established PDAC trained on tens of thousands of non-contrast CT scans~\cite{panda}. {This imbalance reflects the limited public pancreas MRI resources, which have remained small, single-center, and focused mainly on branch-duct IPMNs.} Pancreas segmentation from MRI, which is a prerequisite for any region-of-interest-based classifier, has correspondingly lagged. {Prior MRI-specific segmentation studies typically used fewer than one hundred scans from a single institution, and the largest published MRI pancreas segmentation cohort to date, from our group, contains 767 scans from 499 participants}~\cite{zhang2025large}. {A major bottleneck is therefore a lack of suitable data rather than of algorithms. Without large {multicenter}, multi-vendor MRI resources with risk labels confirmed by histopathology or follow-up, MRI-based IPMN risk stratification could not be developed or externally evaluated at the scale the clinical problem requires.}

Research on pancreatic segmentation has focused mainly on CT imaging, with limited exploration of MRI due to data scarcity~\cite{zhang2023deep}. Cai \textit{et al.} proposed a recurrent neural network-based approach to pancreatic segmentation, achieving a {Dice} coefficient of 76.1\% on 79 T1-weighted (T1W) MRI scans~\cite{cai2017improving}. Salanitri \textit{et al.} introduced a multi-headed decoder structure that generates intermediate segmentation maps and ensembles these predictions to produce the final output. Their method achieved a {Dice} coefficient of 77.5\% on 40 in-house T2-weighted (T2W) MRI scans~\cite{proietto2021hierarchical}. Zhang \textit{et al.} developed {the} 
\textit{PanSegNet} model, which achieved Dice coefficients of 85.0\% on T1W MRI and 86.3\% on T2W MRI~\cite{zhang2025large}. {General-purpose multi-organ models such as \textit{TotalSegmentator MRI}~\cite{dantonoli2025totalsegmentator} and \textit{MRSegmentator}~\cite{haentze2024mrsegmentator} were trained on more scans in total (1,143 and 2,649 CT and MRI examinations, respectively), but these collections span many anatomical structures and body regions, so abdominal MRI with pancreas labels forms only part of them. Such models are not optimized for pancreas MRI; on \textit{Cyst-X}, zero-shot \textit{MRSegmentator} segmented the pancreas less accurately than \textit{PanSegNet} on both sequences (Section~\ref{sec:segmentation results}).}

For IPMN classification, early radiomics-based approaches by Chakraborty \textit{et al.}~\cite{chakraborty2018ct} and Cui \textit{et al.}~\cite{cui2021radiomic} reported areas under the receiver operating characteristic curve (AUCs) up to 88\% using features extracted from manually segmented cysts, with a predominant focus on branch-duct IPMNs. However, main-duct IPMNs, which have a significantly higher risk of malignancy~\cite{corral2019deep,ohtsuka2024international,gonda2024pancreatic}, remain relatively underrepresented. Deep learning methods have evolved from 2D convolutional neural network (CNN)-based feature extractors~\cite{hussein2018deep} to fully end-to-end 3D models such as \textit{DenseINN} and \textit{InceptINN}~\cite{lalonde2019inn}, which adapt pre-trained 2D weights through an ``inflation'' mechanism. Corral \textit{et al.} developed a CNN-based deep learning protocol to classify IPMN malignancy from MRI in a cohort of 139 cases, and reported an AUC of 78\%, comparable to the Fukuoka (77\%) and American Gastroenterological Association (76\%) guidelines~\cite{corral2019deep}. Transformer-based models~\cite{salanitri2022neural} and hybrid frameworks that combine segmentation, radiomics, and deep learning predictions~\cite{yao2023radiomics} have also demonstrated improved classification performance. A recent study from our multicenter collaboration~\cite{bejar} evaluated an overlapping, histopathology-defined subset of 359 T2W MRI examinations from the same seven-center collection, focusing specifically on cyst-level radiomics and fusion strategies for low- versus high-risk IPMN classification. That work established the feasibility of multicenter cyst-focused MRI analysis, but was restricted to T2W imaging and a selected histopathology-defined IPMN cohort. {The present \textit{Cyst-X} study expands this foundation in three ways. It provides the full multicenter benchmark across T1W and T2W MRI, with a separate no-risk population and expert pancreas segmentations. It defines hierarchical three-class and binary classification tasks and evaluates automated segmentation-to-classification, leave-one-center-out, and federated settings. It also releases the data, predefined splits, and models publicly.} {\textit{PanSegNet} and our earlier federated studies provide methodological foundations for this work~\cite{zhang2025large,pan2025ipmn,pan2025adaptive}. The contribution of \textit{Cyst-X} is the consolidated multicenter MRI resource and its common evaluation framework, which distinguishes IPMN-specific risk classification from broader pancreatic phenotype discrimination and characterizes performance across institutions.}

A second, complementary obstacle to multi-institutional MRI-based AI is data governance. Centralizing patient images across jurisdictions is constrained by the Health Insurance Portability and Accountability Act (HIPAA) in the United States, the General Data Protection Regulation in the European Union, and analogous frameworks elsewhere. {These constraints have historically limited multi-institutional studies to narrowly negotiated data-sharing agreements or single-center designs.} Federated learning offers a practical engineering response: model parameters rather than raw images move between sites, and aggregation occurs on a coordinating server. {We treat federation as an applied capability rather than a cryptographic privacy guarantee; formal mechanisms such as differential privacy and secure aggregation are deployment-stage extensions. The empirical question is therefore:} can federated training of an IPMN classifier {across seven institution-partitioned data silos, simulated on a single server,} recover the discrimination achievable when the same images are pooled?

\begin{figure} [htbp]
     \centering\includegraphics[width=0.72\linewidth]{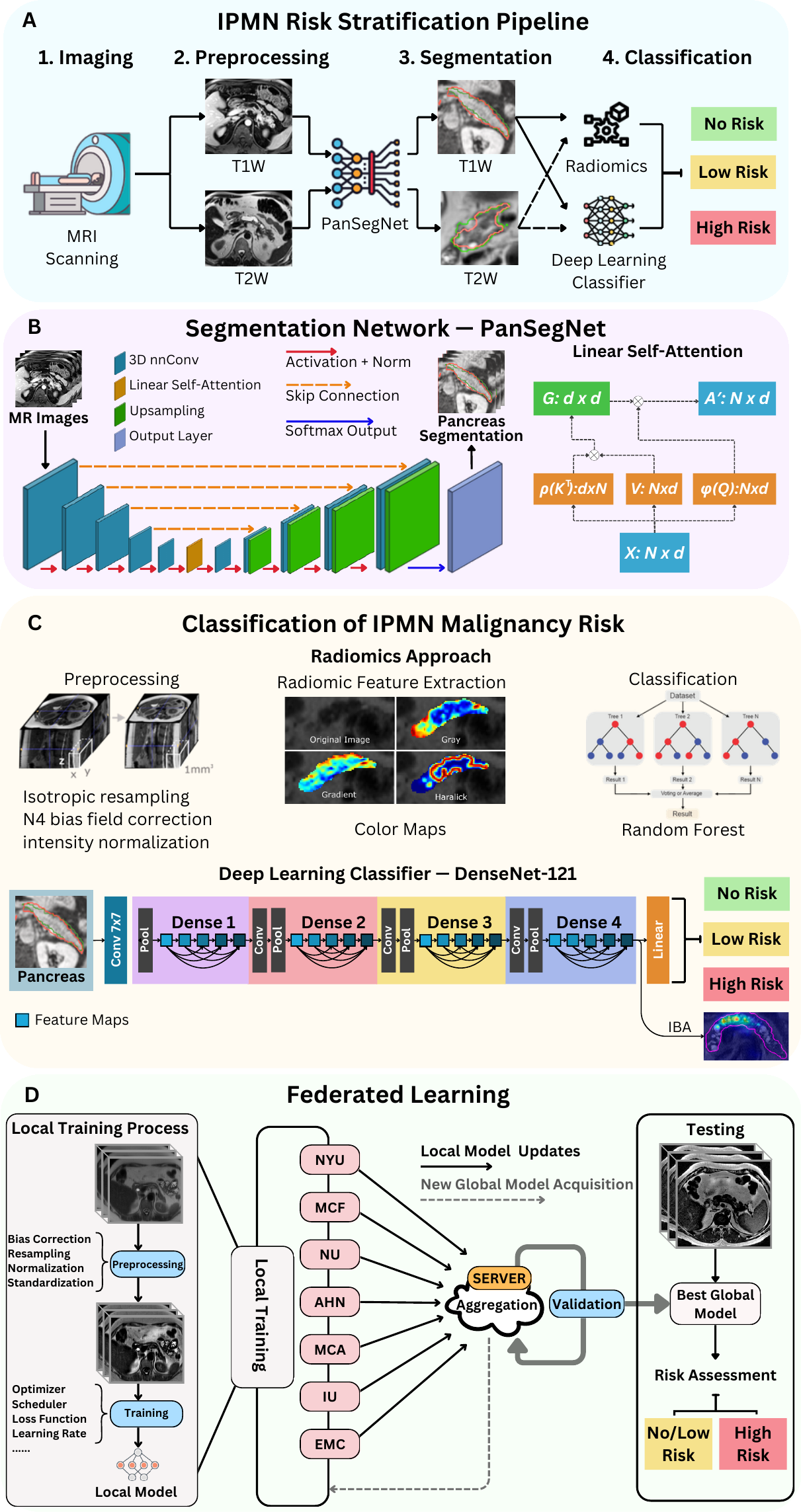}
     \caption{\textit{Cyst-X} Pipeline. {(A) T1W and T2W MRI scans were collected from seven medical centers {(center abbreviations are defined in Section~\ref{sec:cohort})}. (B) {\textit{PanSegNet}, trained and evaluated on \textit{Cyst-X}, segments the pancreas from the MRI scans.} 
     (C) Radiomics analysis and deep learning classification assess the malignancy risk of IPMNs. (D) {Federated optimization is evaluated across seven simulated institutional silos.}}}
     \label{fig: overview}
 \end{figure}

To address all these challenges, we assembled the \textbf{\textit{Cyst-X}} dataset, 1,461 abdominal MRI scans {(723 T1W and 738 T2W)} from 764 patients at seven international centers spanning the United States, Turkey, and the Netherlands, and used it to {benchmark pancreas segmentation and IPMN risk classification, including the effect of automated segmentation on complementary classification tasks} (Fig.~\ref{fig: overview}). {\textit{PanSegNet}, a published \textit{nnU-Net}-based model, provides automated pancreas segmentation. The primary high-risk versus low-risk IPMN analysis uses radiologist-defined pancreas regions of interest (ROIs), while separate experiments evaluate automated ROIs for broad binary and three-class classification on T1W and T2W MRI.} A parallel radiomics pipeline based on 1,409 hand-crafted features, maximum-relevance minimum-redundancy feature selection, and a random forest classifier provides a classical-pipeline baseline against which {a 3D \textit{DenseNet-121} classifier (trained from scratch)} is benchmarked. {We evaluate \textit{DenseNet-121}} under centralized training and under federated training using Federated Averaging (FedAvg)~\cite{mcmahan2017communication} and Federated Proximal optimization (FedProx)~\cite{li2020federated} across {seven simulated} institutional silos.

{The primary contribution of \textit{Cyst-X} is a reusable multicenter benchmark for pancreatic MRI segmentation and risk classification, with shared images, annotations, reference labels, predefined splits, code, and trained models. Radiomics, deep learning, automated segmentation, and federated training provide reference baselines under common evaluation protocols. The principal clinical question is discrimination between high-risk and low-risk IPMN; broad binary and three-class phenotype classification are complementary benchmark tasks. A paired comparison with three blinded radiologists characterizes imaging-only sensitivity--specificity tradeoffs.}

{The experiments identify distinct challenges within this benchmark. \textit{PanSegNet} maintained its Dice scores on held-out centers (86.34\% and 89.26\% on T1W and T2W MRI, versus 86.81\% and 89.62\% internally). For risk classification, IPMN-only AUCs were lower than those for broad binary discrimination. For federated models, internal, pooled external, and center-averaged AUCs diverged, indicating that the comparability of model scores across centers needs to be evaluated alongside discrimination. These analyses provide a reproducible framework for evaluating segmentation, risk classification, and cross-center performance.}

\section{Results}

\subsection{Pancreas segmentation {(\textit{PanSegNet})}}\label{sec:segmentation results}
\begin{figure}[htbp]
\centering
\includegraphics[width=\linewidth]{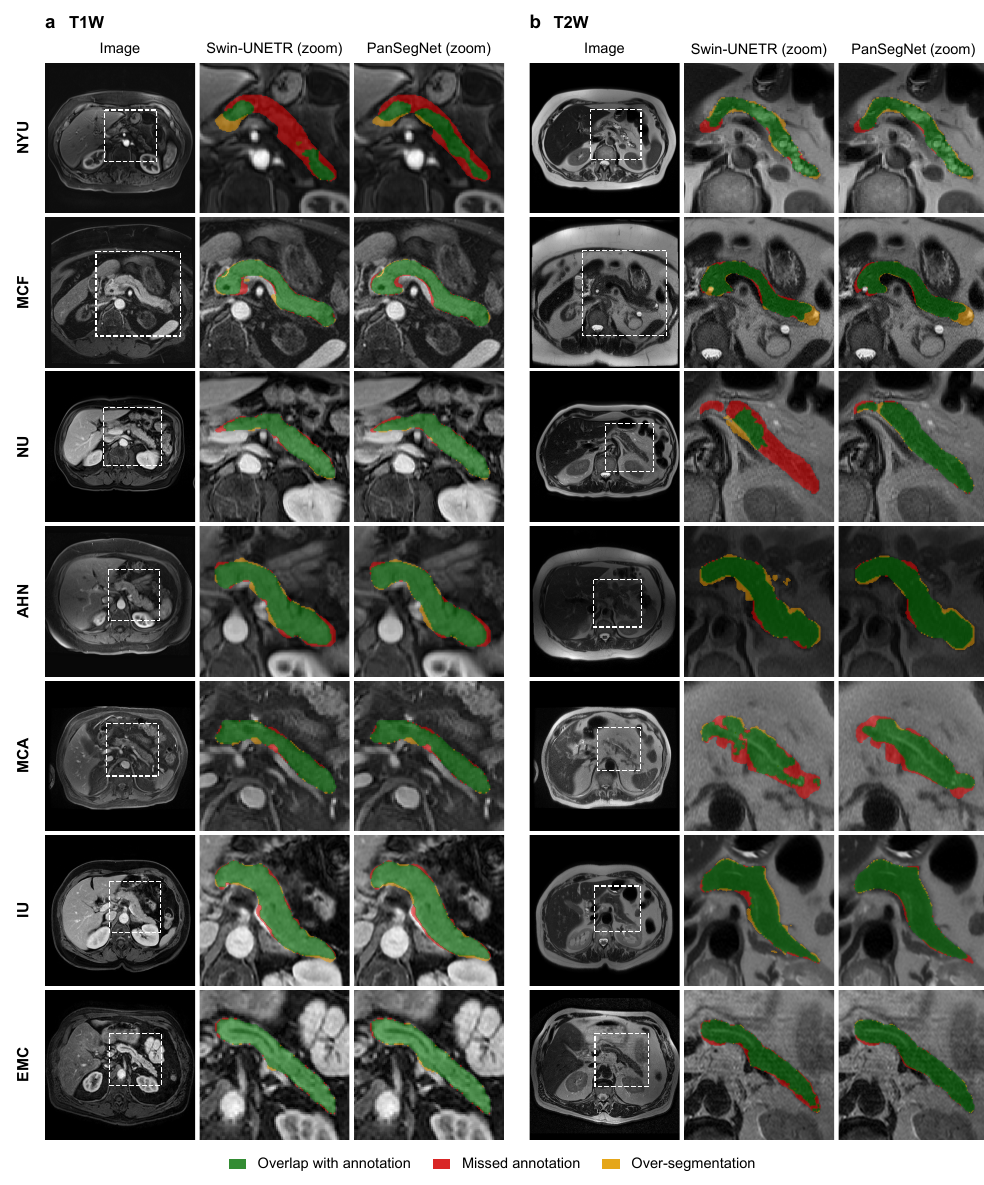}
\caption{\textit{Cyst-X} Segmentation Results. {One representative case per center on (a) T1W and (b) T2W MRI. For each case, the full image is shown with the zoomed region outlined (dashed box), followed by zoomed \textit{Swin-UNETR} and \textit{PanSegNet} predictions. Annotation masks not covered by model predictions are shown in red; model outputs are shown in green (overlap with annotations) and yellow (over-segmentation).}}
\label{fig: segmentation results}
\end{figure}

Accurate pancreas segmentation is a critical prerequisite for precise cyst analysis and classification. Recently, we developed \textit{PanSegNet}~\cite{zhang2025large}, an accurate and efficient segmentation architecture incorporating linear self-attention layers~\cite{shen2021efficient} within the nnU-Net framework~\cite{isensee2021nnu} to enhance global information modeling capabilities while maintaining computational efficiency (Fig.~\ref{fig: overview}). \textit{PanSegNet} achieved mean Dice coefficients of 86.81\% on T1W MRI and 89.62\% on T2W MRI across five-fold cross-validation (Table~\ref{tab: segmentation results}, Fig.~\ref{fig: segmentation results}). The transformer-based \textit{Swin-UNETR} backbone~\cite{hatamizadeh2021swin} reached 79.09\% (T1W) and 76.29\% (T2W) {(zero-shot \textit{MRSegmentator}~\cite{haentze2024mrsegmentator}, 75.53\% and 59.93\%)}. {\textit{PanSegNet}'s mean Dice was therefore higher than that of \textit{Swin-UNETR} by 7.72 percentage points on T1W and 13.33 percentage points on T2W MRI, and the size of this gap varied across centers (Supplementary Tables~\ref{tab: panseg} and~\ref{tab: SwinUnetR}). Because the classifiers operate on the segmented pancreas, Section~\ref{sec:Segmentation quality is a first-class determinant of classification performance} examines how the segmentation source affects downstream classification.}

{Federated training introduces additional implementation considerations for pancreas segmentation.} \textit{PanSegNet} inherits \textit{nnU-Net}'s centralized dataset fingerprinting~\cite{isensee2021nnu}. Federated extensions of nnU-Net that aggregate fingerprints across sites {are now available~\cite{skorupko}; implementing them was outside the scope of this study.} For the federated comparison, we used \textit{Swin-UNETR}, which {can be federated} without modification. {The contribution here is the federated training and evaluation protocol rather than a specific segmentation network: nnU-Net-based federated methods, or any other segmentation model, can replace \textit{Swin-UNETR} within the same protocol.} {The resulting performance loss therefore characterizes this architecture and training protocol rather than segmentation in general.} 
Federated \textit{Swin-UNETR} incurred a Dice penalty of 7.83 percentage points on T1W MRI and 7.10 percentage points on T2W MRI relative to its centralized baseline (Table~\ref{tab: segmentation results}, \textit{FedAvg} rows). {Further controlled experiments are needed to distinguish task-related effects from architecture and optimization effects.}

\begin{table}[tb]
    \centering
    \setlength{\tabcolsep}{4pt}
    \caption{Benchmarking for Segmentation of Pancreas from MRI. {\textit{PanSegNet} values are mean $\pm$ s.d. across patients pooled over the five cross-validation folds; \textit{Swin-UNETR} values (centralized and federated) are mean $\pm$ s.d. of the five fold-level means.}}
    \begin{footnotesize}
    \begin{tabular}{lllllll}
\toprule
\textbf{Method} & \textbf{Dice} & \textbf{Jaccard} & \textbf{Precision} & \textbf{Recall} & \textbf{HD95} & \textbf{ASSD}\\
 & (\%) & (\%) & (\%) & (\%) & (mm) & (mm)\\
\midrule
\multicolumn{7}{l}{\textbf{T1W MRI}}\\
PanSegNet &   86.81$\pm$7.30 &   77.32$\pm$9.74 &   86.56$\pm$8.59 &   87.84$\pm$7.80 & {3.72$\pm$8.86} & {0.74$\pm$1.73} \\
Swin-UNETR & 79.09$\pm$1.40 & 67.19$\pm$1.63 & 79.09$\pm$1.67 & 81.37$\pm$0.74 & 26.55$\pm$8.48 & 7.58$\pm$3.76   \\
+FedAvg&71.26$\pm$2.59 & 58.30$\pm$2.74 & 69.74$\pm$3.09 & 77.13$\pm$1.34 & 43.65$\pm$8.60 & 14.77$\pm$5.01\\
+FedProx($\mu$=0.3)&57.24$\pm$3.41 & 44.16$\pm$2.85 & 56.85$\pm$4.33 & 62.93$\pm$2.80 & 87.27$\pm$8.34 & 31.99$\pm$6.19\\
+FedProx($\mu$=0.01)&68.95$\pm$3.30 & 56.09$\pm$3.37 & 67.33$\pm$3.37 & 75.11$\pm$2.68 & 52.82$\pm$9.97 & 18.76$\pm$5.41\\
+FedProx($\mu$=0.005)&69.77$\pm$2.90 & 56.88$\pm$2.98 & 68.16$\pm$2.98 & 75.82$\pm$2.39 & 49.61$\pm$8.55 & 17.51$\pm$5.39\\
\midrule
\multicolumn{7}{l}{\textbf{T2W MRI}}\\
PanSegNet &  89.62$\pm$6.38 &   81.73$\pm$9.31 &   90.74$\pm$6.45 &   89.11$\pm$8.06 & {2.51$\pm$5.23} & {0.35$\pm$0.65} \\
Swin-UNETR & 76.29$\pm$0.66 & 63.77$\pm$0.82 & 78.70$\pm$2.14 & 76.86$\pm$1.21 & 28.53$\pm$7.01 & 7.57$\pm$2.60  \\
+FedAvg&69.19$\pm$1.51 & 55.48$\pm$1.44 & 70.52$\pm$2.98 & 71.31$\pm$0.95 & 41.87$\pm$5.84 & 11.74$\pm$3.02\\
+FedProx($\mu$=0.3)&58.11$\pm$2.61 & 43.85$\pm$2.22 & 59.01$\pm$4.38 & 61.46$\pm$1.11 & 66.68$\pm$9.10 & 20.87$\pm$4.81\\
+FedProx($\mu$=0.01)&67.25$\pm$1.86 & 53.47$\pm$1.77 & 68.22$\pm$2.99 & 70.23$\pm$1.21 & 48.36$\pm$6.51 & 14.52$\pm$3.64\\
+FedProx($\mu$=0.005)&67.82$\pm$1.44 & 54.16$\pm$1.43 & 68.63$\pm$2.41 & 70.89$\pm$0.84 & 48.64$\pm$5.82 & 13.99$\pm$2.96\\
\bottomrule
    \end{tabular}
    \end{footnotesize}
    \label{tab: segmentation results}
\end{table}

\subsection{IPMN risk-classification benchmark}~\label{sec: results classification}
{The primary clinical analysis distinguishes high-risk from low-risk IPMN. The broad binary and three-class tasks are complementary benchmark analyses:
\begin{itemize}
    \item \textbf{IPMN-only binary classification (primary clinical analysis):} Distinguishes high-risk from low-risk IPMN among patients with established or presumed IPMN. High-risk labels denote histopathology-confirmed high-grade dysplasia or invasive carcinoma, as distinct from Kyoto high-risk stigmata, which are imaging features; low-risk labels are based on histopathology or at least three years of imaging stability.
    \item \textbf{Broad binary classification (complementary benchmark):} Distinguishes high-risk IPMN from low-risk IPMN and the no-risk population, which includes normal pancreas and benign non-IPMN cystic findings. This task assesses broader pancreatic phenotype discrimination.
    \item \textbf{Three-class classification (complementary benchmark):} Separates the no-risk, low-risk IPMN, and high-risk IPMN groups using the same reference definitions. It assesses discrimination among these pancreatic phenotypes rather than risk stratification solely within IPMN.
\end{itemize}
All three tasks use the labeled modality-specific cohorts. Cohort sizes and classification results are reported in Table~\ref{tab: IPMN Classification Results}.}

{Under five-fold cross-validation, DenseNet-121 achieved IPMN-only AUCs of 74.59\% on T1W and 73.45\% on T2W MRI. Broad binary AUCs were 78.67\% and 81.28\%, and three-class macro-averaged AUCs were 73.47\% and 75.76\% {(T1W and T2W, respectively;} Table~\ref{tab: IPMN Classification Results}). The broad task additionally includes patients without IPMN and evaluates discrimination across a wider disease spectrum. Radiologist comparisons use 629 paired labeled cases, with IPMN-only and histopathology-confirmed subsets of 474 and 357 patients, respectively (Table~\ref{tab: ai vs human}).}

\begin{figure}[htbp]
     \centering
     \includegraphics[width=\linewidth]{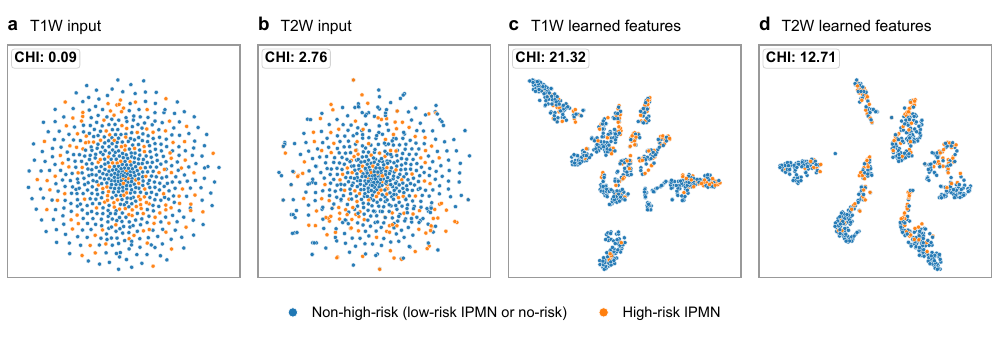}
     \caption{{Broad Binary Classification t-SNE Visualization of (a) T1W and (b) T2W input voxels and of (c) T1W and (d) T2W \textit{DenseNet-121} learned features. While learned representations in (c) and (d) improve clustering and CHI~\cite{calinski1974dendrite} relative to raw inputs in (a) and (b), separation between risk classes remains incomplete. The non-high-risk group comprises low-risk IPMN and the no-risk population.}}\label{fig: tsne}
 \end{figure}

A t-distributed stochastic neighbor embedding (t-SNE) projection of the input voxels and of the \textit{DenseNet-121} penultimate-layer embeddings {of held-out test cases} (Fig.~\ref{fig: tsne}) shows that learned representations partially separate {high-risk IPMN from the non-high-risk group (low-risk IPMN and the no-risk population)}, which the raw images do not. {The Calinski-Harabasz index (CHI) increased from 0.09 to 21.32 on T1W and from 2.76 to 12.71 on T2W MRI. The separation is incomplete,} as high-risk points remain interspersed within the {non-high-risk} region. {This overlap is consistent with the moderate average precision of the centralized \textit{DenseNet-121} classifier (Fig.~\ref{fig: T1W PR},~\ref{fig: T2W PR}).}

\begin{table}[!htbp]
\caption{IPMN Classification Results. Internal AUCs are mean $\pm$ s.d. across the five cross-validation folds, with conditional 95\% CIs from a percentile bootstrap of the fold-mean AUC (test cases resampled within each fold, fitted models held fixed). External results are presented as the pooled AUC across all held-out test cases, with the unweighted mean of the seven center-specific AUCs in parentheses; external 95\% CIs refer to the pooled AUC. The number after FedProx denotes the proximal term coefficient $\mu$. {Unless otherwise indicated, models used radiologist-defined pancreas ROIs. MACs, multiply--accumulate operations; Param., number of parameters. Additional metrics appear in Supplementary Tables~\ref{tab: t1 low vs high classification}--\ref{tab: fusion full}.}}
\begin{center}
{
\footnotesize
\setlength{\tabcolsep}{4.5pt}
\begin{tabular}{lllllll}
\toprule
\textbf{Method} & \textbf{MACs} & \textbf{Param.} & \multicolumn{2}{c}{\textbf{T1W MRI}} & \multicolumn{2}{c}{\textbf{T2W MRI}}\\
\cmidrule(lr){4-5}\cmidrule(lr){6-7}
 & & & AUC (\%) & 95\% CI (\%) & AUC (\%) & 95\% CI (\%)\\
\midrule
\multicolumn{7}{l}{\textbf{IPMN-only binary: high-risk versus low-risk IPMN}}\\
\multicolumn{7}{l}{\footnotesize\textcolor{gray}{Internal 5-fold cross-validation: T1W $n$=488 (149/339), T2W $n$=496 (152/344)}}\\
3D Radiomics & -- & -- & 73.19$\pm$2.55 & [68.21, 78.30] & 73.64$\pm$2.91 & [68.93, 78.18] \\
DenseNet-121 & 18.31G & 11.24M & 74.59$\pm$3.28 & [69.38, 79.76] & 73.45$\pm$5.28 & [68.89, 78.03] \\
\addlinespace[2pt]
\multicolumn{7}{l}{\footnotesize\textcolor{gray}{External leave-one-center-out evaluation: T1W $n$=488, T2W $n$=496}}\\
3D Radiomics & -- & -- & 64.56 (65.61) & [59.21, 70.06] & 64.66 (61.77) & [58.96, 69.79] \\
DenseNet-121 & & & 66.93 (64.13) & [61.81, 72.00] & 69.30 (67.00) & [64.35, 73.82] \\
\midrule
\multicolumn{7}{l}{\textbf{Broad binary: high-risk IPMN versus non-high-risk (low-risk IPMN + no-risk)$^{\dagger}$}}\\
\multicolumn{7}{l}{\footnotesize\textcolor{gray}{Internal 5-fold cross-validation: T1W $n$=652 (149/339/164), T2W $n$=655 (152/344/159)}}\\
3D Radiomics & -- & -- & 78.02$\pm$0.97 & [73.45, 82.06] & 78.91$\pm$1.71 & [74.20, 82.86] \\
ResNet-34 & 183.04G & 63.47M & 76.72$\pm$1.65 & [72.16, 80.89] & 79.44$\pm$2.73 & [75.26, 83.54] \\
ResNet-50 & 138.07G & 46.17M & 77.83$\pm$2.50 & [73.48, 81.69] & 78.96$\pm$2.82 & [74.80, 82.91] \\
EfficientNet-B0 & 1.21G & 4.69M & 76.33$\pm$5.34 & [71.69, 80.62] & 77.10$\pm$2.10 & [72.38, 81.06] \\
DenseNet-121 & 18.31G & 11.24M & 78.67$\pm$4.02 & [74.41, 82.50] & 81.28$\pm$2.44 & [77.28, 85.22] \\
+FedAvg & & & 77.99$\pm$4.88 & [73.46, 82.11] & 76.00$\pm$3.70 & [71.88, 80.10] \\
+FedProx(0.1) & & & 79.28$\pm$5.15 & [74.85, 83.14] & 77.51$\pm$3.62 & [73.42, 81.55] \\
+FedProx(0.3) & & & 78.86$\pm$4.41 & [74.62, 82.76] & 79.59$\pm$3.00 & [75.56, 83.34] \\
\multicolumn{3}{l}{DenseNet-121 (PanSegNet ROIs)} & 77.40$\pm$4.80 & [73.24, 81.34] & 76.36$\pm$2.77 & [71.87, 80.28] \\
\multicolumn{3}{l}{DenseNet-121 (Swin-UNETR ROIs)} & 70.48$\pm$4.26 & [65.72, 75.02] & 72.11$\pm$2.23 & [67.06, 76.77] \\
\addlinespace[2pt]
\multicolumn{7}{l}{\footnotesize\textcolor{gray}{External leave-one-center-out evaluation: T1W $n$=652, T2W $n$=655}}\\
3D Radiomics & -- & -- & 67.82 (68.28) & [62.69, 72.72] & 71.53 (66.48) & [66.67, 75.75] \\
DenseNet-121 & & & 71.35 (74.22) & [66.49, 76.20] & 77.49 (71.18) & [73.17, 81.44] \\
+FedAvg & & & 66.90 (73.12) & [62.28, 71.58] & 65.15 (72.28) & [60.39, 69.71] \\
+FedProx(0.1) & & & 62.54 (73.80) & [57.65, 67.46] & 74.18 (68.61) & [69.78, 78.37] \\
+FedProx(0.3) & & & 65.89 (72.09) & [61.40, 70.32] & 62.35 (71.00) & [58.04, 66.83] \\
\midrule
\multicolumn{7}{l}{\textbf{Three-class: high-risk IPMN versus low-risk IPMN versus no-risk$^{\dagger}$}}\\
\multicolumn{7}{l}{\footnotesize\textcolor{gray}{Internal 5-fold cross-validation: T1W $n$=652 (149/339/164), T2W $n$=655 (152/344/159)}}\\
ResNet-34 & 183.04G & 63.47M & 72.38$\pm$4.97 & [69.31, 75.43] & 74.07$\pm$1.73 & [71.10, 77.07] \\
ResNet-50 & 138.07G & 46.17M & 69.85$\pm$3.42 & [66.61, 72.90] & 75.13$\pm$0.75 & [72.13, 78.05] \\
EfficientNet-B0 & 1.21G & 4.69M & 67.54$\pm$3.34 & [64.29, 70.76] & 73.31$\pm$3.82 & [70.27, 76.31] \\
DenseNet-121 & 18.31G & 11.25M & 73.47$\pm$4.49 & [70.48, 76.71] & 75.76$\pm$1.29 & [72.42, 78.72] \\
\multicolumn{3}{l}{DenseNet-121 (PanSegNet ROIs)} & 71.25$\pm$3.36 & [67.95, 74.42] & 72.11$\pm$1.53 & [68.98, 75.06] \\
\multicolumn{3}{l}{DenseNet-121 (Swin-UNETR ROIs)} & 64.74$\pm$3.16 & [61.38, 68.19] & 69.00$\pm$2.14 & [65.65, 72.18] \\
\bottomrule
\end{tabular}}
\footnotetext{\parbox[t]{\linewidth}{\raggedright{Numbers in parentheses are high-risk/low-risk/no-risk patients. $^{\dagger}$Includes the no-risk group (patients without cysts or with non-IPMN benign cysts); these AUCs reflect broader phenotype discrimination and are not a direct measure of high-risk versus low-risk IPMN discrimination.}}}
\label{tab: IPMN Classification Results}
\end{center}
\end{table}

{We generated} \textit{Grad-CAM}~\cite{selvaraju2017grad} and Information Bottleneck Attribution (\textit{IBA})~\cite{schulz2020restricting,demir2021information} {maps to show the image regions associated with each prediction} (Supplementary Fig.~\ref{fig: iba}). {These maps indicate association with the prediction rather than causal explanation.}


\subsubsection{{High-risk versus low-risk IPMN classification}}\label{sec: High-risk versus low-risk IPMN classification}
{To isolate malignancy-risk stratification from the broader distinction between high-risk IPMN and normal or non-IPMN pancreatic phenotypes, {we retrained the models after excluding the no-risk group from the training, validation, and test subsets. Numerical decision thresholds were computed using only the corresponding IPMN-only validation subsets, without access to test labels; the 90\% specificity target itself was chosen after observing the radiologists' specificities on the same patients (Supplementary Section~\ref{sec:Clinically meaningful threshold calibration})}. This IPMN-only analysis directly addresses the clinical question encountered after an IPMN has been identified: whether MRI can distinguish lesions with high-grade dysplasia or invasive malignancy from low-risk IPMNs.}
{Under five-fold cross-validation, \textit{DenseNet-121} achieved mean AUCs of 74.59$\pm$3.28\% on T1W MRI (95\% CI, 69.38\%--79.76\%) and 73.45$\pm$5.28\% on T2W MRI (95\% CI, 68.89\%--78.03\%) (Table~\ref{tab: IPMN Classification Results}). The corresponding AUCs for 3D radiomics were 73.19$\pm$2.55\% (95\% CI, 68.21\%--78.30\%) and 73.64$\pm$2.91\% (95\% CI, 68.93\%--78.18\%), respectively. The two approaches showed similar mean discrimination, with overlapping confidence intervals on both modalities. Compared with the broad binary task, the mean AUC for \textit{DenseNet-121} was lower by 4.08 percentage points on T1W and 7.83 percentage points on T2W MRI. {Stratifying risk within IPMN was therefore harder than separating high-risk IPMN from the broader non-high-risk population.}} {Under leave-one-center-out evaluation, \textit{DenseNet-121} achieved pooled AUCs of 66.93\% on T1W MRI (95\% CI, 61.81\%--72.00\%) and 69.30\% on T2W MRI (95\% CI, 64.35\%--73.82\%). The corresponding unweighted center-averaged AUCs were 64.13\% and 67.00\%, respectively. Center-level results are reported in Supplementary Tables~\ref{tab: t1 low vs high classification} and~\ref{tab: t2 low vs high classification}.}

{High-risk prevalence differed across centers, from 11--12\% at NU to 58--70\% at MCA (Supplementary Table~\ref{tab: classification distribution}). To test whether discrimination depended on these differences, we fitted logistic regression models with center as a fixed effect; the \textit{DenseNet-121} score remained independently associated with high-risk IPMN (odds ratio per s.d., 2.31 on T1W and 1.91 on T2W MRI under internal cross-validation, and 1.85 and 2.08 under leave-one-center-out evaluation). The leave-one-center-out AUCs above were obtained on held-out centers whose prevalence was not available during training.} {Mean average precision across folds was 54.81--60.67\% for \textit{DenseNet-121} and 58.14--59.37\% for 3D radiomics, approximately twice the high-risk prevalence of about 31\% (Fig.~\ref{fig: roc}a--d).} {Raw \textit{DenseNet-121} probabilities were miscalibrated, with internal calibration slopes of 0.42 on T1W and 0.19 on T2W MRI. After each fold model was recalibrated on its own validation predictions, the slopes were 0.59 and 0.74, the calibration intercepts were \mbox{$-0.19$} on both modalities, and the Brier scores (0.189 and 0.186) were lower than those of a prevalence-only prediction (0.212 and 0.213). At threshold probabilities of 0.20--0.40, the recalibrated models had a higher net benefit than treating all or no patients; under leave-one-center-out evaluation, this held at a threshold of 0.30 on both modalities (Supplementary Fig.~\ref{fig: calibration} and Supplementary Tables~\ref{tab:zw-calibration} and~\ref{tab:zw-netbenefit}).}

\subsubsection{{Broad high-risk versus non-high-risk classification}}
{The broad binary task distinguished high-risk IPMN from a combined non-high-risk group (low-risk IPMN and the no-risk population). Under five-fold cross-validation, \textit{DenseNet-121} achieved mean AUCs of 78.67$\pm$4.02\% on T1W MRI (95\% CI, 74.41\%--82.50\%) and 81.28$\pm$2.44\% on T2W MRI (95\% CI, 77.28\%--85.22\%), compared with 78.02$\pm$0.97\% and 78.91$\pm$1.71\%, respectively, for 3D radiomics (Table~\ref{tab: IPMN Classification Results}). Among the centralized deep-learning backbones evaluated, \textit{DenseNet-121} had the highest mean AUC on both modalities. Its T2W AUC exceeded that of radiomics by 2.37 percentage points, with overlapping 95\% CIs. Accuracy, sensitivity, specificity, and external evaluation results are reported in Supplementary Tables~\ref{tab: t1 classification} and~\ref{tab: t2 classification}.} {Because the comparator group includes patients without IPMN, these results characterize broad high-risk discrimination rather than malignancy-risk stratification solely among patients with IPMN.}

\subsubsection{{Three-class pancreatic phenotype classification}}
{We also evaluated discrimination among the no-risk, low-risk IPMN, and high-risk IPMN groups using five-fold cross-validation. Three-class AUC was calculated as the macro-average of the one-versus-rest AUCs. \textit{DenseNet-121} achieved the highest mean AUC among the evaluated backbones on both T1W MRI (73.47$\pm$4.49\%; 95\% CI, 70.48\%--76.71\%) and T2W MRI (75.76$\pm$1.29\%; 95\% CI, 72.42\%--78.72\%) (Table~\ref{tab: IPMN Classification Results}). The corresponding T1W and T2W AUCs were 72.38\% and 74.07\% for \textit{ResNet-34}, 69.85\% and 75.13\% for \textit{ResNet-50}, and 67.54\% and 73.31\% for \textit{EfficientNet-B0}. These results provide a benchmark for distinguishing the three pancreatic phenotypes.} {Because the no-risk group includes patients without cysts, the macro-averaged AUC captures broader phenotype discrimination and should not be interpreted as a direct measure of high-risk versus low-risk IPMN discrimination.}

\subsubsection{{Effect of segmentation source on classification performance}}\label{sec:Segmentation quality is a first-class determinant of classification performance}
To assess the importance of accurate pancreas segmentation in classification, we evaluated how different ROI sources affect \textit{DenseNet-121}’s performance. Specifically, we compared classification results using ROIs generated by \textit{PanSegNet} and \textit{Swin-UNETR}, both under centralized learning, against a baseline using radiologist-defined ROIs. For every classification test case, the automated pancreas mask was generated using a segmentation model that had not been trained on that patient. As shown in Table~\ref{tab: IPMN Classification Results}, 
{using \textit{PanSegNet}'s masks resulted in a smaller decline in AUC than using \textit{Swin-UNETR}'s masks, consistent with its higher segmentation accuracy.}
{For 3-class classification, the mean AUC dropped from {73.47\%} (radiologist ROI) to {71.25\%} with \textit{PanSegNet}, and further to {64.74\%} with \textit{Swin-UNETR} on T1W MRI. On T2W MRI, the AUC declined from {75.76\%} to {72.11\%} (\textit{PanSegNet}) and to {69.00\%} (\textit{Swin-UNETR}). A similar trend was observed in broad binary classification: on T1W, AUC dropped from {78.67\%} to {77.40\%} (\textit{PanSegNet}) and {70.48\%} (\textit{Swin-UNETR}); on T2W, from {81.28\%} to {76.36\%} and {72.11\%}, respectively.} {These results demonstrate the effect of segmentation source on broad binary and three-class classification. Automated-ROI performance was not evaluated for the primary high-risk versus low-risk IPMN task.}



\subsubsection{Comparison against blinded radiologists on a {paired} 629-case subset}\label{sec:Comparison with expert radiologists}

\begin{table}[p]
\vspace*{-58pt}
\caption{Comparison of radiologist, radiomics, and deep-learning performance for pancreatic cyst risk stratification. Three blinded abdominal radiologists assessed all 629 cases with paired T1W/T2W MRI under imaging-only conditions; their assessments were also analyzed in the IPMN-only cohort (n=474) and histopathology-confirmed IPMN subset (n=357).}
\setlength{\tabcolsep}{3.5pt}
\begin{center}
\begin{scriptsize}\renewcommand{\arraystretch}{0.84}
{
\begin{tabular}{llllll}
\toprule
\textbf{Method} & \textbf{Input} & \multicolumn{2}{c}{\textbf{Internal}} & \multicolumn{2}{c}{\textbf{External}}\\
\cmidrule(lr){3-4}\cmidrule(lr){5-6}
 & & Sens & Spec & Sens & Spec\\
\midrule
\multicolumn{6}{l}{\textbf{629 broad cases (high-risk versus non-high-risk (low-risk+no-risk))}}\\
Radiologist 1&T1W+T2W&64.08&89.94&{--}&{--}\\
Radiologist 2&T1W+T2W&32.39&97.13&{--}&{--}\\
Radiologist 3&T1W+T2W&41.55&94.66&{--}&{--}\\
\bf{Average}&\bf{T1W+T2W}&\bf{46.01}&\bf{93.91}&{--}&{--}\\
\addlinespace[1pt]
3D Radiomics&T1W Only&40.85\,(24.65)&90.35\,(96.92)&47.18\,(13.38)&83.78\,(94.66)\\
3D Radiomics&T2W Only&45.07\,(26.06)&90.14\,(96.30)&30.28\,(16.90)&92.61\,(95.28)\\
\addlinespace[1pt]
DenseNet-121&T1W Only&48.59\,(54.23)&88.50\,(80.70)&48.59\,(58.45)&87.47\,(78.64)\\
DenseNet-121&T2W Only&57.04\,(35.21)&88.50\,(91.58)&49.30\,(59.15)&87.06\,(82.96)\\
Early Fusion&T1W+T2W&45.77\,(64.79)&90.76\,(77.00)&57.04\,(62.68)&82.14\,(78.44)\\
Late Fusion&T1W+T2W&36.62\,(16.20)&89.73\,(97.54)&52.82\,(23.94)&85.63\,(92.40)\\
Siamese Fusion&T1W+T2W&47.18\,(47.89)&88.71\,(89.94)&42.25\,(23.24)&84.39\,(86.65)\\
Logit Fusion&T1W+T2W&45.07\,(54.93)&89.12\,(79.88)&44.37\,(63.38)&84.19\,(74.74)\\
\midrule
\multicolumn{6}{l}{\textbf{474 IPMN-only cases (low-risk versus high-risk)}}\\
Radiologist 1&T1W+T2W&64.08&86.45&{--}&{--}\\
Radiologist 2&T1W+T2W&32.39&96.08&{--}&{--}\\
Radiologist 3&T1W+T2W&41.55&92.47&{--}&{--}\\
\bf{Average}&\bf{T1W+T2W}&\bf{46.01}&\bf{91.67}&{--}&{--}\\
\addlinespace[1pt]
3D Radiomics&T1W Only&45.07\,(34.51)&86.45\,(90.06)&27.46\,(24.65)&90.96\,(87.95)\\
3D Radiomics&T2W Only&39.44\,(31.69)&88.86\,(94.28)&31.69\,(16.20)&93.07\,(92.47)\\
\addlinespace[1pt]
DenseNet-121&T1W Only&45.77\,(51.41)&86.14\,(78.92)&36.62\,(47.18)&86.75\,(75.00)\\
DenseNet-121&T2W Only &41.55\,(42.25)&86.75\,(83.73)&47.89\,(58.45)&88.86\,(70.48)\\
Early Fusion&T1W+T2W&45.07\,(42.96)&87.05\,(88.86)&31.69\,(61.97)&90.06\,(53.61)\\
Late Fusion&T1W+T2W&42.25\,(54.93)&85.24\,(72.29)&39.44\,(35.92)&86.45\,(88.55)\\
Siamese Fusion&T1W+T2W&40.14\,(38.73)&84.94\,(84.04)&34.51\,(57.75)&84.94\,(76.51)\\
Logit Fusion&T1W+T2W&40.85\,(18.31)&84.04\,(96.39)&33.80\,(58.45)&86.75\,(48.80)\\
\midrule
\multicolumn{6}{l}{\textbf{357 histopathology-confirmed IPMN cases (histopathology low-risk versus high-risk)}}\\
Radiologist 1&T1W+T2W&64.08&79.53&{--}&{--}\\
Radiologist 2&T1W+T2W&32.39&93.95&{--}&{--}\\
Radiologist 3&T1W+T2W&41.55&88.37&{--}&{--}\\
\bf{Average}&\bf{T1W+T2W}&\bf{46.01}&\bf{87.29}&{--}&{--}\\
\addlinespace[1pt]
3D Radiomics&T1W Only&21.83\,(43.66)&90.23\,(78.14)&17.61\,(38.03)&89.30\,(70.70)\\
3D Radiomics&T2W Only&33.10\,(42.25)&84.65\,(81.40)&9.86\,(27.46)&96.28\,(84.65)\\
\addlinespace[1pt]
DenseNet-121&T1W Only&26.76\,(43.66)&86.51\,(71.63)&23.94\,(45.77)&90.23\,(69.30)\\
DenseNet-121&T2W Only&30.28\,(72.54)&89.77\,(36.74)&64.08\,(73.94)&71.16\,(45.12)\\
Early Fusion&T1W+T2W&26.76\,(61.27)&86.98\,(50.23)&30.99\,(29.58)&87.91\,(82.33)\\
Late Fusion&T1W+T2W&35.92\,(63.38)&82.33\,(43.26)&28.87\,(51.41)&86.51\,(68.37)\\
Siamese Fusion&T1W+T2W&19.01\,(33.10)&87.91\,(75.81)&18.31\,(30.99)&91.63\,(89.30)\\
Logit Fusion&T1W+T2W&28.17\,(73.24)&90.23\,(40.00)&23.24\,(53.52)&91.16\,(73.49)\\
\bottomrule
\end{tabular}}
\footnotetext{\parbox[t]{\textwidth}{\scriptsize\raggedright {Model rows in each block come from models trained for that task: broad binary models (629-case block), IPMN-only models (474-case block), and models retrained on histopathology-confirmed cases (357-case block).} AI and radiomics models underwent patient-level cross-validation (internal testing) and leave-one-center-out evaluation (external testing). Sensitivity (Sens) measures high-risk IPMN detection; specificity (Spec) measures identification of non-high-risk cases in the {629-case broad cohort} and low-risk IPMN in the IPMN-only analyses. Single-modality and T1W-T2W fusion models are indicated. Model inputs for radiologist comparisons used radiologist-defined pancreas regions of interest to isolate classification performance from automated segmentation errors. {Values outside parentheses use validation-selected thresholds}; parenthetical values use a 0.50 probability threshold. {Comparisons are exploratory and describe performance at the reported operating points rather than at matched sensitivity or specificity. Radiologist averages are arithmetic means across the three participating readers, not consensus predictions.} {Radiologist sensitivities are identical across the three cohorts because all three contain the same 142 high-risk patients. Specificity denominators are 487, 332, and 215 patients in the three blocks, respectively.} All values are percentages. {--, not applicable: each radiologist scored every case once, so internal and external evaluation do not apply.}}}
\label{tab: ai vs human}
\end{scriptsize}
\end{center}
\end{table}

{To characterize model and radiologist performance on the same patients, we conducted an exploratory imaging-only comparison using the 629 paired cases described in Section~\ref{sec:Radiologist visual scoring} rather than the full 764-patient cohort.
The remaining 135 patients were excluded because they lacked paired T1W and T2W sequences or ground-truth risk labels. On this benchmark subset, three abdominal radiologists scored cases using Kyoto criteria imaging features while blinded to histopathology, clinical history, and prior imaging (Table~\ref{tab: ai vs human}). To evaluate classification capability independent of automated segmentation error, the model inputs for this comparison were cropped using radiologist-defined ROIs.}
{Radiologist sensitivity for high-risk IPMN varied widely: 64.08\%, 32.39\%, and 41.55\% for Radiologists 1, 2, and 3,} with corresponding specificities of 89.94\%, 97.13\%, and 94.66\%, and a mean radiologist sensitivity of 46.01\% at a mean specificity of 93.91\%.
{At validation-selected operating thresholds (Supplementary Section~\ref{sec:Clinically meaningful threshold calibration}), T2W DenseNet-121 models {trained separately for each task and evaluated by internal cross-validation} showed different sensitivity--specificity tradeoffs across cohorts (Table~\ref{tab: ai vs human}).}

{In the broad 629-case cohort, sensitivity was 57.04\% and specificity was 88.50\%, compared with radiologist averages of 46.01\% and 93.91\%, respectively. Thus, higher sensitivity relative to the radiologist average came at lower specificity; Radiologist 1 exceeded the model in both sensitivity (64.08\%) and specificity (89.94\%). In the 474-case IPMN-only cohort, model sensitivity and specificity were 41.55\% and 86.75\%, both below the corresponding radiologist averages of 46.01\% and 91.67\%. In the 357-case histopathology-confirmed IPMN subset, model sensitivity was lower (30.28\% versus 46.01\%), whereas specificity was higher (89.77\% versus 87.29\%).}
{At the fixed 0.50 threshold, the broad-cohort T2W classifier reached 35.21\% sensitivity and 91.58\% specificity. Modality fusion did not consistently improve on T2W alone; for example, Siamese fusion reached 47.18\% sensitivity and 88.71\% specificity at the validation-selected threshold.}
{Conditional 95\% confidence intervals for the internal T2W \textit{DenseNet-121} comparisons are reported in Supplementary Table~\ref{tab:zw-reader-ci}, and paired differences between the model and the mean across radiologists are reported in Supplementary Table~\ref{tab:zw-paired-differences}. The patient-level bootstrap procedure and its scope are described in Supplementary Section~\ref{sec:zw-paired-analysis}.}


{These comparisons characterize cohort-dependent sensitivity--specificity tradeoffs under imaging-only conditions and do not establish overall model superiority over radiologists. The higher sensitivity of the broad-task T2W model relative to the radiologist mean was accompanied by lower specificity, and this sensitivity advantage was not observed in the IPMN-only or histopathology-confirmed analyses. The radiologist mean summarizes the three participating readers rather than a consensus assessment or an estimate of performance across the wider radiologist population. Because clinical history, laboratory findings and prior imaging were withheld, these results do not evaluate performance against full guideline-based clinical management.}

\subsubsection{{Histopathology-confirmed-only sensitivity analysis}}
{When histopathologic confirmation was unavailable, low-risk labels were defined by longitudinal imaging stability. Such labels may introduce incorporation bias, because imaging information then contributes to both the label and the model input. To examine performance in the histopathology-confirmed population, we excluded the 117 follow-up-defined patients (Table~\ref{tab: dataset composition}). All models were retrained from scratch on this histopathology-confirmed cohort. Internal cross-validation and leave-one-center-out fold assignments were kept unchanged, with the follow-up-defined cases removed from each fold. On this subset, the internal single-modality DenseNet-121 on T2W MRI achieved a sensitivity of 30.28\% at a specificity of 89.77\% (Table~\ref{tab: ai vs human}). Sensitivity was lower than the radiologist average of 46.01\%, whereas specificity was higher than the radiologist average of 87.29\% in the same histopathology-confirmed subset.}

{This analysis uses histopathology-confirmed cases for both training and evaluation, yielding a smaller cohort with a different case mix and a histopathologic reference standard. We also restricted the held-out predictions of the original models, with their original validation-selected thresholds, to the same 357 histopathology-confirmed patients (Supplementary Table~\ref{tab:zw-frozen-pathology}). Internally, the broad-task T2W model achieved 57.04\% sensitivity, 78.60\% specificity, and 67.26\% AUC on this subset; the histopathology-trained T2W model achieved 30.28\%, 89.77\%, and 62.35\%, respectively. On the same 357 patients, the original IPMN-only models achieved AUCs of 66.51\% on T1W and 62.15\% on T2W MRI, lower than their pooled AUCs in the full IPMN-only cohort (70.84\% and 67.86\%; Supplementary Tables~\ref{tab:zw-frozen-pathology} and~\ref{tab:zw-calibration}), consistent with both the different case mix of this subset and the possibility that surveillance-defined low-risk lesions are easier to distinguish.}

\subsection{{Federated classification performance depends on modality and evaluation summary}\label{sec:FL}}
\begin{figure}[tb]
\centering
\captionsetup[subfloat]{justification=centering}
{\raggedright\small\textbf{IPMN-only task: high-risk versus low-risk IPMN}\par}
\subfloat[T1W ROC]{\includegraphics[width=.24\linewidth]{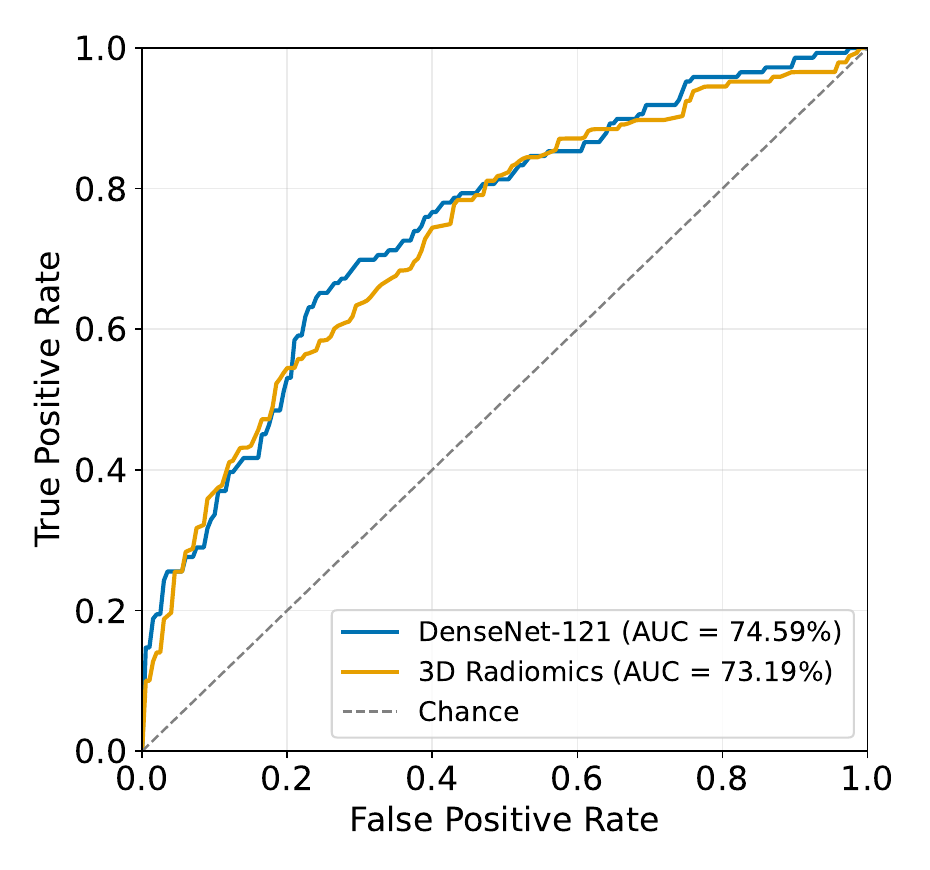}}\hfill
\subfloat[T2W ROC]{\includegraphics[width=.24\linewidth]{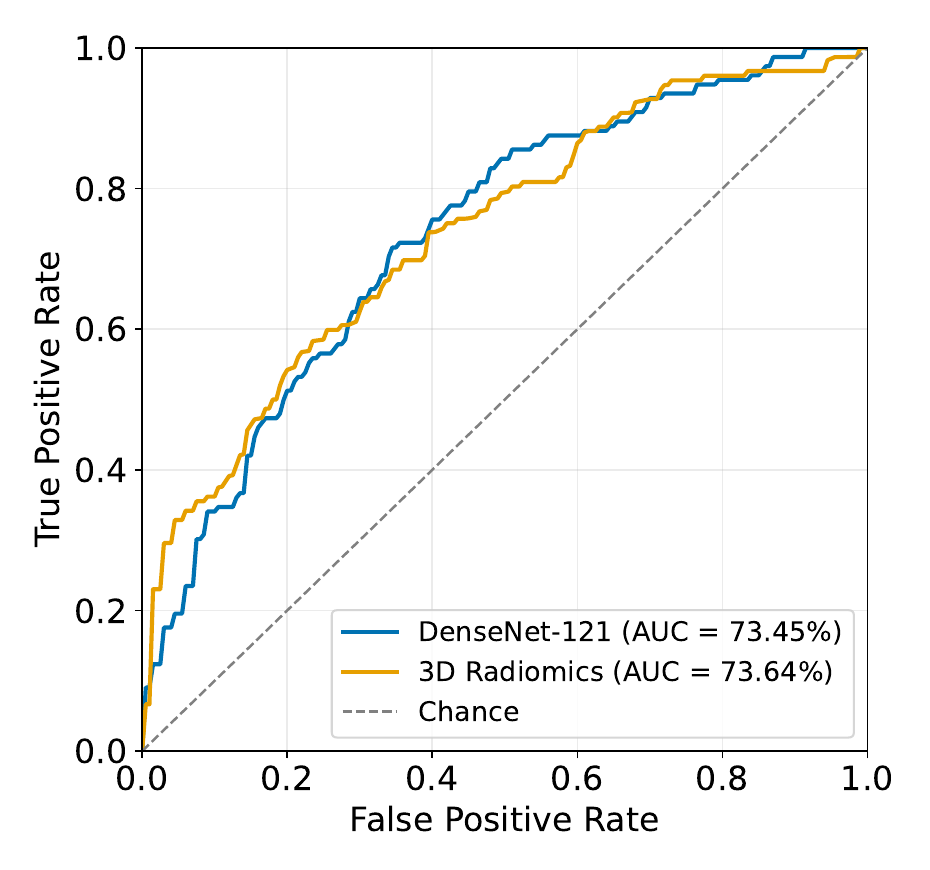}}\hfill
\subfloat[T1W PR]{\includegraphics[width=.24\linewidth]{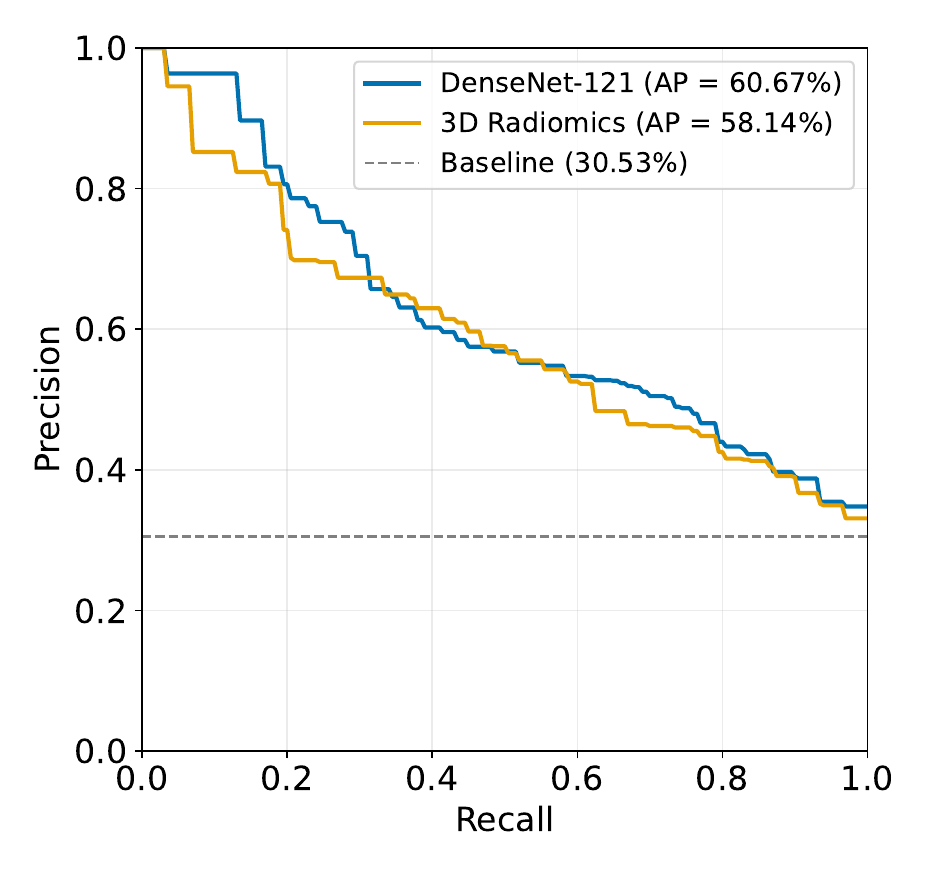}}\hfill
\subfloat[T2W PR]{\includegraphics[width=.24\linewidth]{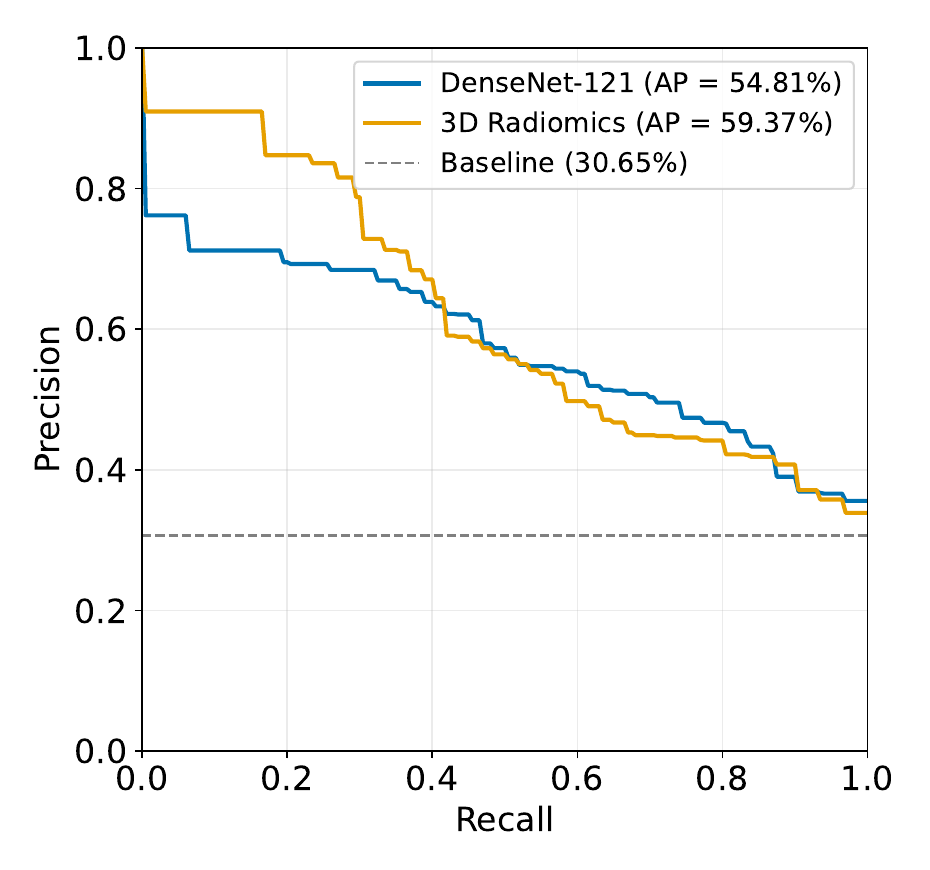}}

\vspace{2mm}
{\raggedright\small\textbf{Broad binary task: high-risk IPMN versus non-high-risk}\par}
\subfloat[T1W ROC]{\includegraphics[width=.24\linewidth]{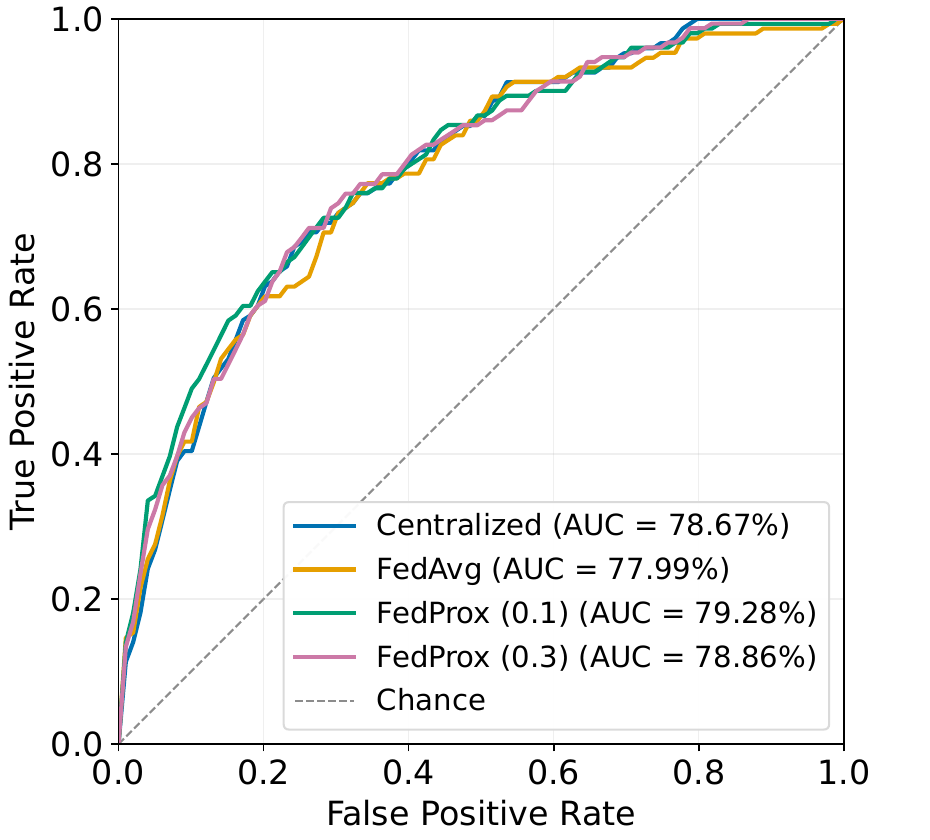}}\hfill
\subfloat[T2W ROC]{\includegraphics[width=.24\linewidth]{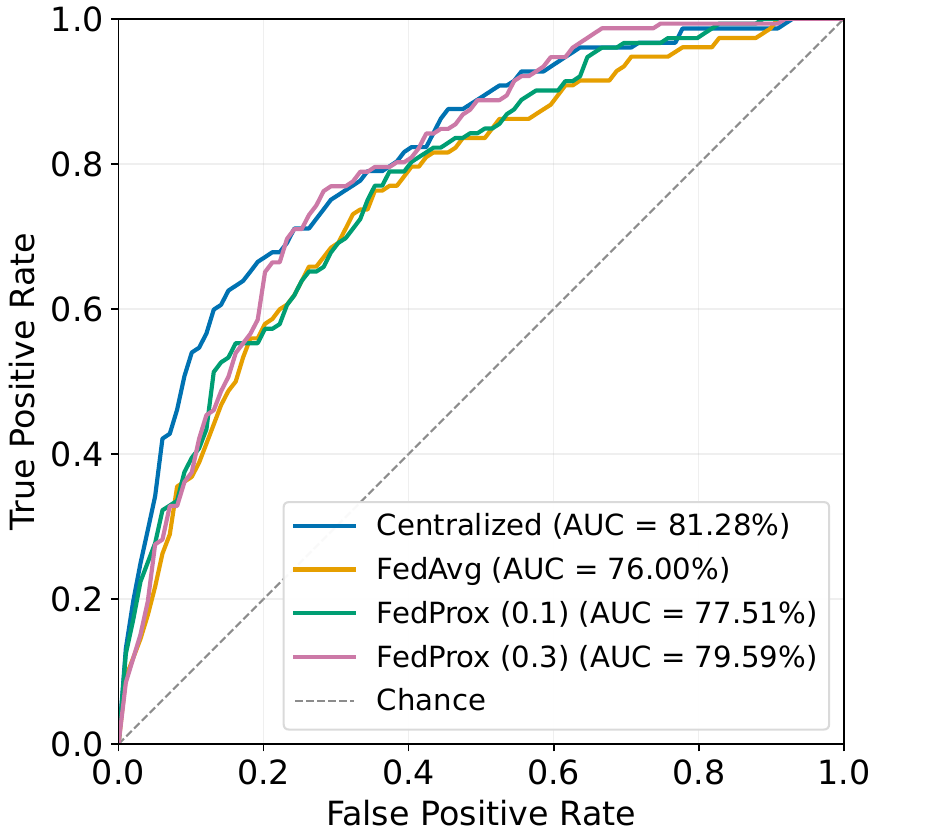}}\hfill
\subfloat[T1W PR\label{fig: T1W PR}]{\includegraphics[width=.24\linewidth]{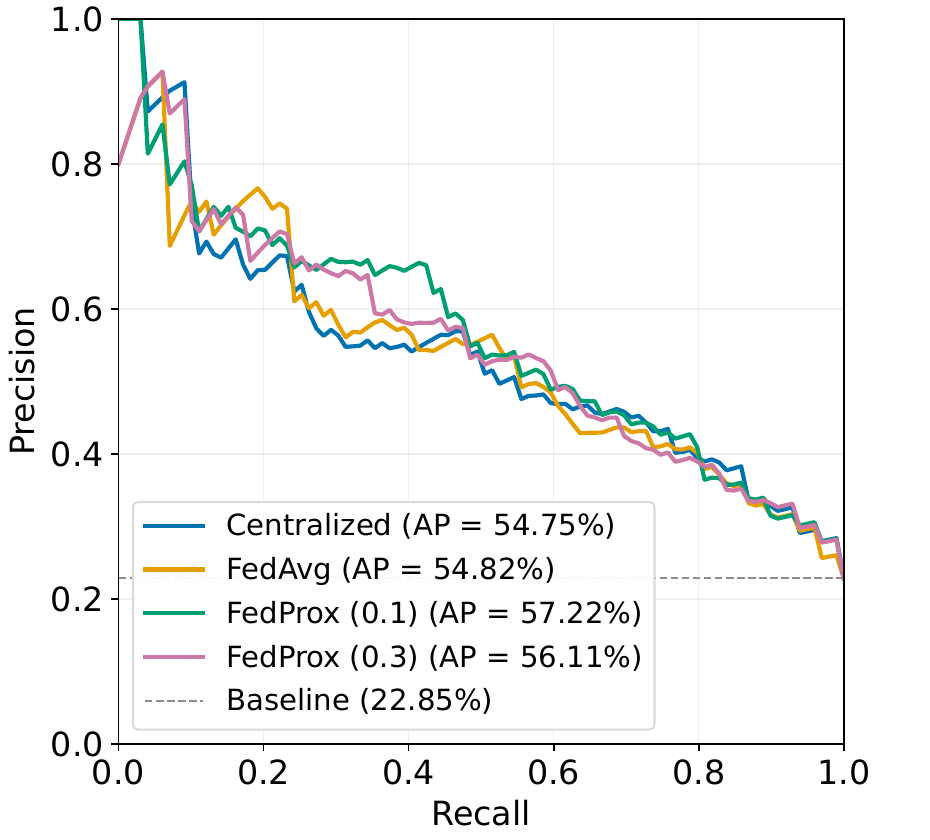}}\hfill
\subfloat[T2W PR\label{fig: T2W PR}]{\includegraphics[width=.24\linewidth]{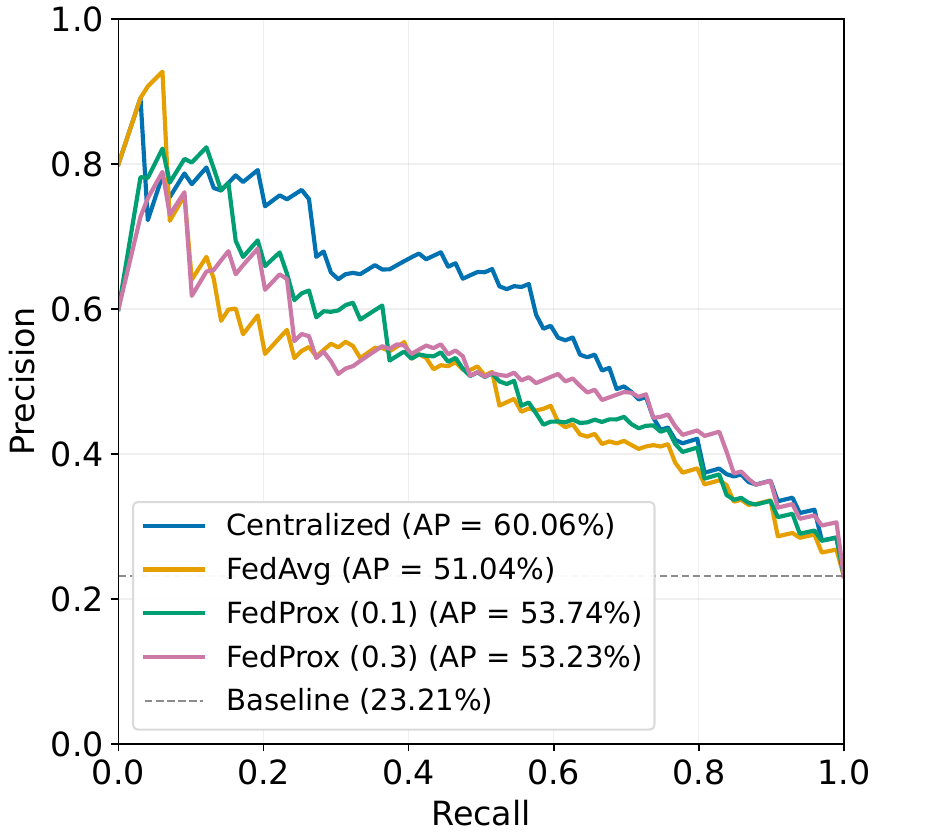}}
\caption{{Mean receiver operating characteristic (ROC) and precision--recall (PR) curves across five internal cross-validation folds. (a)--(d) Primary IPMN-only task (high-risk versus low-risk IPMN), comparing DenseNet-121 and 3D radiomics. (e)--(h) Broad binary task (high-risk versus non-high-risk), comparing centralized and federated DenseNet-121. Legends give fold-mean AUC and average precision (AP); the dashed PR line marks the prevalence of the high-risk class. Plotted PR curves average interpolated precision across folds, whereas AP is computed per fold from the step-wise curve and then averaged.}}
\label{fig: roc}
\end{figure}

{We tested whether off-the-shelf federation algorithms preserve classification discrimination across seven heterogeneous, multi-vendor MRI centers. Using \textit{FedAvg}~\cite{mcmahan2017communication} and \textit{FedProx}~\cite{li2020federated}, we trained models on institution-partitioned data silos simulated on a centralized server (Section~\ref{sec:FLimplementation}). On T1W MRI, \textit{FedProx} with $\mu=0.1$ reached a mean AUC of 79.28\% across five folds, compared with 78.67\% under centralized training. On T2W MRI, this configuration reached 77.51\%, compared with 81.28\% under centralized training; the best federated T2W result was 79.59\% with $\mu=0.3$ (Table~\ref{tab: IPMN Classification Results}). Selected federated configurations approached centralized performance when assessed by internal fold-mean AUC. This finding did not extend consistently to pooled predictions: for T2W FedProx with $\mu=0.3$, pooled internal AUC was 65.72\%, compared with 78.17\% for centralized training, despite fold-mean AUCs of 79.59\% and 81.28\%, respectively (Supplementary Table~\ref{tab:zw-matched-auc}). Federation therefore did not uniformly preserve classification discrimination across evaluation summaries. The ROC and PR curves in Fig.~\ref{fig: roc}e--h show centralized average precision of 54.75\% on T1W and 60.06\% on T2W MRI, compared with chance baselines of 22.85\% and 23.21\%, respectively.}

{For segmentation, by contrast, federated \textit{Swin-UNETR} incurred substantial Dice penalties (Table~\ref{tab: segmentation results}). The mean Dice score dropped from 79.09\% for centralized \textit{Swin-UNETR} to 71.26\% under \textit{FedAvg} on T1W MRI, and from 76.29\% to 69.19\% on T2W MRI (Supplementary Table~\ref{tab: SwinUnetRFedAvg}). Evaluation metrics degraded further under \textit{FedProx} (Supplementary Tables~\ref{tab: SwinUnetRFedProx03},~\ref{tab: SwinUnetRFedProx001} and~\ref{tab: SwinUnetRFedProx0005}). In these experiments, federated training degraded segmentation substantially more than classification assessed by internal fold-mean AUC. One practical configuration supported by the present experiments is to distribute a centrally trained segmentation model while federating the downstream classifier.}

\subsection{External evaluation: leave-one-center-out cross-validation}\label{sec:external}
{To approximate the most clinically realistic setting, in which a model is applied at a site whose data were never seen during training, we performed leave-one-center-out cross-validation. Models were trained on six centers and tested on the seventh, with each center held out once.}

Segmentation generalized well {across unseen institutions. Under leave-one-center-out evaluation, \textit{PanSegNet} achieved global mean Dice scores of 86.34\% (T1W) and 89.26\% (T2W), close to its internal results of 86.81\% (T1W) and 89.62\% (T2W) (Supplementary Table~\ref{tab:panseg external}).} 

{For broad binary classification, DenseNet-121 reached pooled external AUCs of 71.35\% on T1W and 77.49\% on T2W MRI. Internal fold-mean AUCs were 78.67\% and 81.28\%, respectively. When both settings were summarized by pooled AUC, the internal-to-external difference was smaller (74.94\% versus 71.35\% on T1W and 78.17\% versus 77.49\% on T2W; Supplementary Table~\ref{tab:zw-matched-auc}). The unweighted center-averaged external AUCs were 74.22\% on T1W and 71.18\% on T2W MRI. Center-level external AUCs ranged from 55.00\% to 84.25\% on T1W and from 50.89\% to 87.50\% on T2W MRI (Supplementary Tables~\ref{tab: t1 classification} and~\ref{tab: t2 classification}). These results demonstrate substantial variation in discrimination across held-out centers.}

{External leave-one-center-out evaluation revealed differences between pooled and center-averaged AUCs (Supplementary Tables~\ref{tab: t1 classification}, \ref{tab: t2 classification} and~\ref{tab:zw-matched-auc}). On T2W MRI, centralized pooled AUC was 77.49\%, compared with 65.15\% for FedAvg, 74.18\% for FedProx ($\mu=0.1$), and 62.35\% for FedProx ($\mu=0.3$); the corresponding center-averaged AUCs were 71.18\%, 72.28\%, 68.61\%, and 71.00\%. On T1W MRI, centralized pooled AUC was 71.35\%, compared with 66.90\%, 62.54\%, and 65.89\% for the same federated configurations; center-averaged AUCs were 74.22\%, 73.12\%, 73.80\%, and 72.09\%, respectively. Thus, every federated configuration had a lower pooled AUC than centralized training (by 3.31--15.14 percentage points; all 95\% CIs excluded zero), whereas center-averaged AUCs differed from centralized training by at most 2.57 percentage points (all 95\% CIs included zero). These differences are consistent with model scores that are scaled differently across centers and with differences in cohort composition. {Recalibration supported this interpretation. When each federated model was recalibrated using only its own validation predictions, without data from the held-out center, pooled external AUC rose from 62.35\% to 78.17\% for T2W FedProx (\mbox{$\mu=0.3$}) and from 65.15\% to 72.05\% for T2W FedAvg. After recalibration, the pooled external AUCs of the federated configurations differed from those of recalibrated centralized training (76.99\% on T1W and 75.56\% on T2W MRI) by \mbox{$-5.32$} to \mbox{$+2.61$} percentage points, compared with \mbox{$-15.14$} to \mbox{$-3.31$} percentage points before recalibration (Supplementary Table~\ref{tab:zw-calibration}).} {Under external evaluation, T1W sensitivity at the fixed 0.5 threshold} was 55.70\% under centralized training, compared with 18.79\%, 24.83\%, and 62.42\% under FedAvg, FedProx ($\mu=0.1$), and FedProx ($\mu=0.3$), respectively.}

{The variation across held-out centers motivates prospective validation across scanners and patient populations, together with evaluation of adaptation to local data (Section~\ref{sec:discussion}).}



\subsection{Inter-{center} heterogeneity and its imaging correlates}\label{sec:Inter-center heterogeneity and its imaging correlates}
{

{Fig.~\ref{fig: MRQy} presents the analysis of 21 image-quality indicators computed with \textit{MRQy} and projected into two dimensions with uniform manifold approximation and projection (\textit{UMAP}).} In the T1W projections {(rows 1--2)}, some clusters are composed of centers with similar {slice thickness}. One mixed cluster consists of the NU and IU centers (green and brown, respectively), both of which have a mean {slice thickness} of 3.3 mm. Other isolated clusters are composed of EMC (pink), MCF (orange), and NYU (blue) centers, with mean {slice thicknesses} of 2.5 mm, 4.9 mm, and 5.3 mm, respectively. Similarly, in the T2W projections, one mixed cluster consists of the NU and NYU centers (green and blue), with mean {slice thicknesses} of 5.6 mm and 5.3 mm, respectively. Other isolated clusters are composed of the EMC (pink) and MCF (orange) centers, with mean {slice thicknesses} of 6.8 mm and 4.9 mm, respectively. 
{These observations suggest that slice thickness contributes to inter-center imaging heterogeneity, alongside other acquisition factors.}


\begin{figure}[htbp]
\centering
\includegraphics[width=\linewidth]{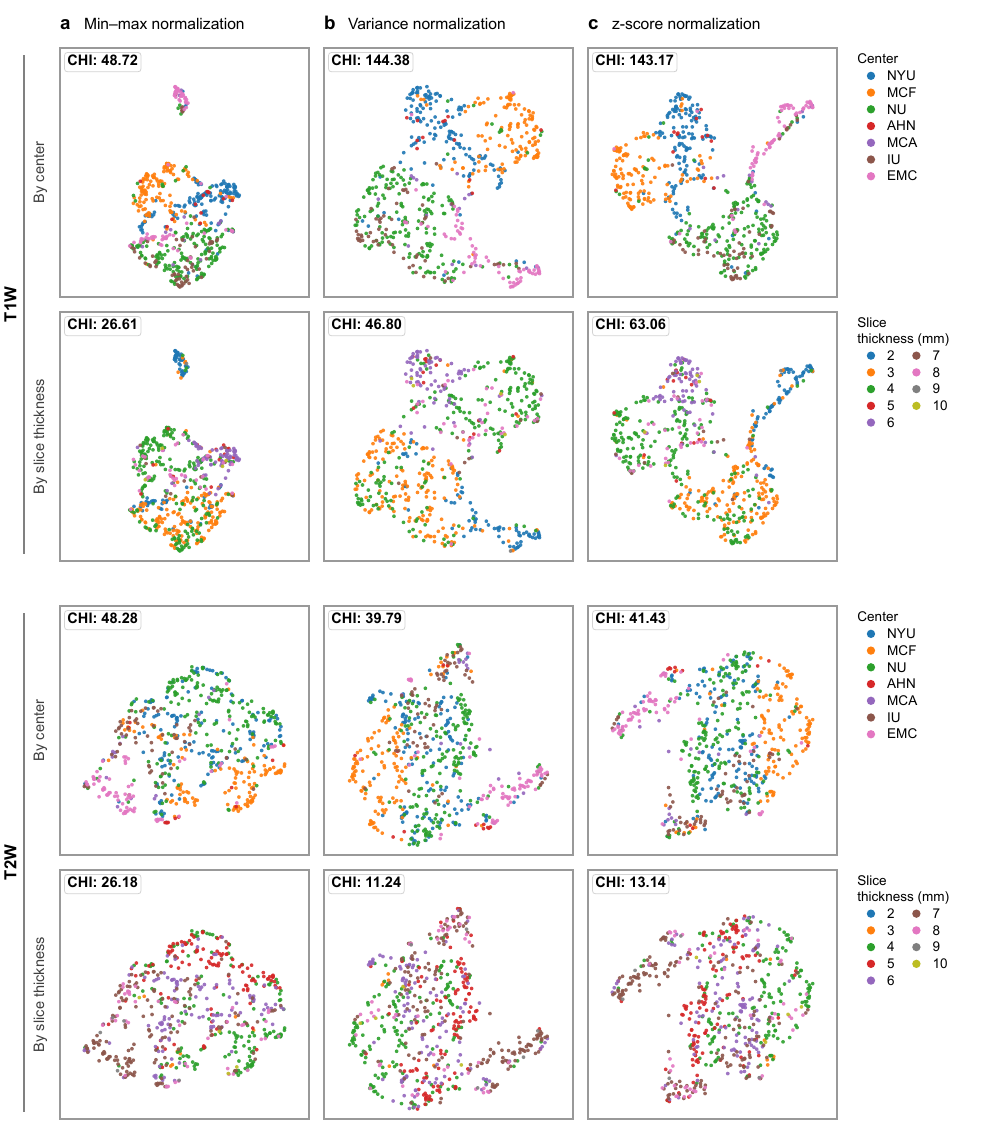}
\caption{\textit{MRQy} analysis. {UMAP projections of the 21 image-quality metrics after (a) min--max, (b) variance and (c) z-score normalization. Rows 1--2: T1W scans; rows 3--4: T2W scans. Points are colored by center (rows 1 and 3) or by slice thickness (rows 2 and 4). CHI, Calinski--Harabasz index~\cite{calinski1974dendrite} of the groups shown by the coloring; higher values indicate better-separated groups. UMAP axes are dimensionless and are not shown.}}
\label{fig: MRQy}
\end{figure}

\section{Discussion\label{sec:discussion}}
{\textit{Cyst-X} provides a common basis for reproducible evaluation of pancreatic MRI segmentation and risk classification across institutions. By sharing images, annotations, labels, splits, and reference models, it exposes how performance differs between anatomical segmentation, clinically focused risk classification, and broader phenotype discrimination.}

{\textit{PanSegNet}'s mean Dice was similar in internal cross-validation and on held-out centers: 86.81\% versus 86.34\% on T1W and 89.62\% versus 89.26\% on T2W MRI. For risk classification, the primary clinical IPMN-only analysis yielded internal AUCs of 74.59\% and 73.45\%, below those of the broad binary task. Federated evaluation also identified the comparability of model scores across centers as a target for methodological development.}

{The no-risk population complements, rather than replaces, IPMN-only risk stratification. The broad high-risk versus non-high-risk task tests whether high-risk IPMN exhibits an imaging phenotype distinct from low-risk IPMN, benign or non-IPMN cystic findings, and normal pancreas. The IPMN-only analysis isolates the clinically specific question of malignancy-risk stratification once an IPMN has been identified, and the three-class benchmark tests whether these pancreatic states are separable, rather than assuming that they lie on a single binary spectrum. In the IPMN-only analysis, DenseNet-121 achieved internal AUCs of 74.59\% on T1W and 73.45\% on T2W MRI (Table~\ref{tab: IPMN Classification Results}), compared with 78.67\% and 81.28\% in the broader binary task. Discrimination decreased once the no-risk population was excluded, so the clinically distinct tasks must be reported separately. For context, Corral \textit{et al.} reported an AUC of 78\% for MRI-based IPMN classification in a single-center cohort of 139 patients~\cite{corral2019deep}; the present estimates come from a seven-center cohort with additional leave-one-center-out evaluation.}

IPMN management is longitudinal and multifactorial; hence, it requires sequential imaging, growth kinetics, evolving morphologic features, patient comorbidities, and multidisciplinary assessment. Accordingly, a single-time point MRI-based risk estimate should be interpreted as a complementary input rather than a stand-alone determinant of surgical decision-making. {The reader comparison provides a clinical reference for interpreting the benchmark, with sensitivity--specificity tradeoffs that varied across cohorts. Higher specificity was accompanied by lower sensitivity in some readers, but the study did not assess their decision rationale or the clinical consequences of these operating points. The model specificity target was selected after observing reader performance, although numerical thresholds were subsequently determined using validation data alone. The comparisons are therefore exploratory, and the reported confidence intervals describe patient-sampling uncertainty conditional on the fitted models, thresholds, centers and three participating readers. Future studies with prospectively specified operating criteria and radiologist-with-AI assessments are needed to determine whether model assistance improves clinical decisions.}


{Federated optimization provided a complementary benchmark of classification performance across institutions. Although selected configurations approached centralized internal fold-mean AUC, pooled performance was substantially lower for some federated models. For T2W FedProx with $\mu=0.3$, pooled internal and external AUCs were 65.72\% and 62.35\%, compared with 78.17\% and 77.49\% for centralized training, whereas center-averaged external AUCs were similar (71.00\% and 71.18\%; Supplementary Table~\ref{tab:zw-matched-auc}). Interpretation requires accounting for the evaluation design: pooled internal predictions combine outputs from different cross-validation models, whereas pooled external predictions combine outputs from models trained with different centers excluded. Differences among fold-mean, pooled and center-averaged AUCs may therefore reflect variation in model score scales, cohort composition and within-center discrimination; validation-based recalibration removed most of the pooled-AUC deficit of the federated models (Section~\ref{sec:external}), indicating that differences in score scale across models, rather than a loss of within-center discrimination, account for much of this deficit. Because raw \textit{DenseNet-121} probabilities were miscalibrated, recalibration on local validation data should accompany any use of these scores as probabilities. The results support reporting these complementary summaries together rather than concluding that federation generally preserves performance. Segmentation exhibited a larger performance reduction under the evaluated federated Swin-UNETR protocol; however, because the segmentation and classification experiments used different architectures and optimization settings, this difference should not be interpreted as an intrinsic limitation of federated segmentation. Federated nnU-Net variants~\cite{skorupko} can be substituted for Swin-UNETR within the same benchmark protocol. Finally, the present experiments simulate institutional silos and do not incorporate formal privacy mechanisms such as secure aggregation~\cite{bonawitz2017practical} or differential privacy~\cite{geyer2017differentially,cheng2022differentially}; prospective multi-server deployment with explicit calibration and privacy safeguards remains an important direction for future work.}

{Several limitations bound the conclusions of this study. First, the lower performance and center-level variability observed under external leave-one-center-out evaluation reflect sensitivity to local acquisition protocols and cohort distributions. Because both disease prevalence and imaging characteristics vary substantially across institutions, site identity is a potential confounder; therefore, the observed cross-center performance differences cannot be attributed solely to biological variation. {Nevertheless, after adjustment for center, the \textit{DenseNet-121} score remained independently associated with high-risk IPMN, both under internal cross-validation (odds ratio per s.d., 2.31 [95\% CI, 1.79--2.98] on T1W and 1.91 [1.49--2.45] on T2W MRI) and under leave-one-center-out evaluation (1.85 [1.44--2.37] and 2.08 [1.60--2.71]).} Second, the reference standard combines histopathology from resection or diagnostic tissue sampling with surveillance-defined low-risk status. These sources provide different levels of diagnostic certainty: sampled tissue may underrepresent the highest lesion grade, whereas imaging stability does not establish histologic grade. The histopathology-only analyses remove surveillance-defined labels but do not eliminate uncertainty related to tissue sampling or selection for intervention. Individual follow-up durations were unavailable beyond the minimum three-year inclusion criterion, and incomplete procedure-level metadata prevented a separate resection-confirmed analysis. Third, the histopathology-confirmed low-risk group is susceptible to spectrum bias. Many of these lesions underwent intervention after showing concerning preoperative imaging features, so they may not fully represent the broader population of indolent IPMNs managed by surveillance.}

{The primary IPMN-only analysis and radiologist comparisons used radiologist-defined pancreas ROIs and therefore assess classification performance with manual localization. Validation of the complete automated pipeline for IPMN-only risk classification, with both segmentation and classification developed independently of the held-out center, remains necessary before drawing conclusions about automated cross-institutional performance. \textit{Cyst-X} is a retrospective benchmark rather than a system validated for clinical deployment. A prospective study applying a locked classifier to consecutive incident IPMN cases at a new site, with radiologist assessment captured before model inference, would provide a more direct estimate of deployment performance. Finally, the cohort spans a twenty-year acquisition period (2004–2024), during which scanner hardware, sequence parameters, and reconstruction software evolved substantially. Temporal heterogeneity and longitudinal drift could not be comprehensively evaluated because of anonymization constraints across institutions; dedicated temporal validation will therefore be required before prospective clinical deployment.}

{\textit{Cyst-X} establishes a reusable benchmark for pancreatic MRI segmentation and risk classification, with shared data, annotations, evaluation splits, and reference baselines. The clinically focused IPMN-only analysis, complementary phenotype tasks, radiologist comparisons, and matched internal--external summaries make the strengths and limitations of current models explicit. Independent groups have already begun to reuse the resource~\cite{sgourakis}. Because images, labels, splits and models are released together, future methods, including those that integrate clinical information and cyst-fluid genomics~\cite{singhi2026pancreaseq}, can be compared on the same patients under the same protocols.}

\section{Methods}
\subsection{The \textit{Cyst-X} cohort}\label{sec:cohort}
{The study was approved by the Northwestern University Institutional Review Board (\#STU00214545); participating institutions obtained local ethics approval or established reliance agreements as required by their institutional and national regulations. The requirement for informed consent was waived for retrospective analysis where permitted by the approving ethics committees.} {The \textit{Cyst-X} resource comprises 1,461 sequence volumes (723 T1W and 738 T2W) from 764 patients, of whom 697 had both sequences available. Risk-labeled classification sets contained 652 T1W and 655 T2W cases. The paired, risk-labeled radiologist-comparison cohort comprised 629 patients, including 474 with IPMN; 357 of these had histopathologic confirmation and 117 had surveillance-defined low-risk labels. These denominators distinguish the full imaging resource from the task-specific analysis cohorts.} Patients aged 18 years or older were imaged at seven international medical centers between March 2004 and June 2024: New York University (NYU) Langone Health, Mayo Clinic Florida (MCF), Northwestern University (NU), Allegheny Health Network (AHN), Mayo Clinic Arizona (MCA), Istanbul University Faculty of Medicine (IU), and Erasmus Medical Center (EMC). {All MRI scans were de-identified at the contributing centers according to institutional protocols before transfer to our center.} Scanners spanned three principal vendors (GE, Siemens, and Philips) and both 1.5-Tesla and 3-Tesla field strengths; demographic characteristics across centers are summarized in Table~\ref{tab: dataset composition}. 
{Cases eligible for risk classification were assigned to high-risk IPMN, low-risk IPMN, or no-risk pancreas. Low-risk labels were established through either histopathology or longitudinal surveillance, and the categories were defined as follows:}
\begin{enumerate}
\item \textbf{High-risk}:  IPMN with high-grade dysplasia or worse, including carcinoma in situ and invasive carcinoma, confirmed by endoscopic-ultrasound-guided biopsy or surgical histopathology. 
{\item \textbf{Low-risk, histopathology-confirmed}: IPMN with low- or intermediate-grade dysplasia confirmed on resection or diagnostic tissue sampling.}
\item {\textbf{Low-risk, surveillance-defined}: IPMN lesions without biopsy that remained stable over at least three years of imaging follow-up, with stability defined as growth of less than 2.5 millimeters per year and absence of worrisome features or high-risk stigmata as defined by the Kyoto criteria.}


\item {\textbf{No-risk pancreas}}: {Individuals with either normal pancreatic imaging without cystic lesions or benign cysts not associated with IPMN, as determined on imaging.}

\end{enumerate}

{For histopathology-confirmed cases, index MRI examinations were acquired within six months before the confirmatory procedure. Histopathology-based labels reflected the grade documented in the available surgical or diagnostic tissue specimen; diagnostic sampling may not capture the highest grade present within a lesion. Counts stratified by surgical resection versus diagnostic tissue sampling and details of discrepancy adjudication were unavailable for most centers. Surveillance-defined low-risk labels indicate stability under the specified imaging criteria rather than histologic confirmation of low-grade disease.}


\begin{table}[p]
\setlength{\tabcolsep}{2.5pt}\small
\centering
\caption{\textit{Cyst-X} dataset composition. Data distribution {and histopathology versus imaging-follow-up counts are shown below. ``Low'' and ``High'' stand for low-risk and high-risk IPMN. The diagram shows how clinical endpoints map to risk labels and to the three classification tasks.}}
\begin{tabular}{@{}>{\raggedright\arraybackslash}p{2.3cm}*{7}{>{\raggedright\arraybackslash}p{1.2cm}}@{}}
\toprule
\textbf{Center} & \textbf{NYU} & \textbf{MCF} & \textbf{NU} & \textbf{AHN} & \textbf{MCA} & \textbf{IU} & \textbf{EMC}\\
\midrule
Imaging device & Siemens, GE & Siemens, GE & Siemens & N/A & Siemens, GE & Siemens, Philips & Siemens, GE\\
MRI magnet (T) & 1.5, 3 & 1.5, 3 & 1.5, 3 & N/A & 1.5, 3 & {1.5} & {1.5}\\
\midrule
\multicolumn{8}{@{}l}{\textbf{T1W cohort}}\\
Patient count & 162 & 148 & 206 & 17 & 25 & 74 & 91\\
Female & 98 & 81 & 109 & 12 & 14 & 42 & 43\\
Male & 64 & 67 & 97 & 5 & 11 & 32 & 48\\
Mean age (y) & 63.0 & 65.0 & 63.9 & N/A & 71.1 & 64.1 & 50.9\\
\midrule
\multicolumn{8}{@{}l}{\textbf{T2W cohort}}\\
Patient count & 162 & 143 & 207 & 27 & 23 & 73 & 103\\
Female & 97 & 78 & 109 & 19 & 10 & 41 & 50\\
Male & 65 & 65 & 98 & 8 & 13 & 32 & 53\\
Mean age (y) & 62.9 & 65.4 & 63.9 & N/A & 70.1 & 63.8 & 52.2\\
\bottomrule
\end{tabular}\label{tab: dataset composition}

\vspace{6pt}
{
\begin{tabular}{@{}lccccccccc@{}}
\toprule
 & \multicolumn{3}{c}{\textbf{T1W}} & \multicolumn{3}{c}{\textbf{T2W}} & \multicolumn{3}{c}{\textbf{Paired}}\\
\cmidrule(lr){2-4}\cmidrule(lr){5-7}\cmidrule(lr){8-10}
\textbf{Reference standard} & Low & High & Total & Low & High & Total & Low & High & Total\\
\midrule
Histopathology & 222 & 149 & 371 & 227 & 152 & 379 & 215 & 142 & 357\\
$\ge$3-year imaging follow-up & 117 & 0 & 117 & 117 & 0 & 117 & 117 & 0 & 117\\
\midrule
\textbf{Total} & 339 & 149 & 488 & 344 & 152 & 496 & 332 & 142 & 474\\
\bottomrule
\end{tabular}}

\vspace{4pt}
   \centering \includegraphics[width=0.84\linewidth]{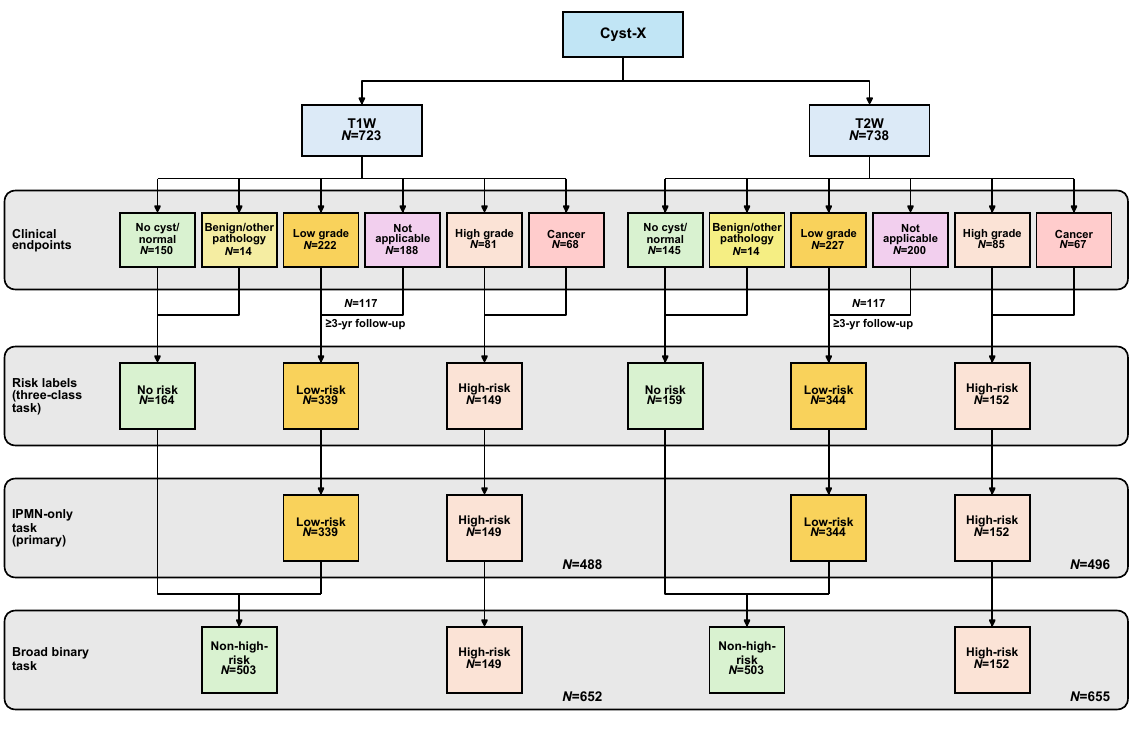}
\parbox{\textwidth}{\footnotesize\raggedright{Note: Ground-truth labels were assigned using a hierarchical reference standard. When histopathologic confirmation was available, histopathology was considered the definitive reference standard. In the absence of histopathologic confirmation, longitudinal imaging follow-up demonstrating at least three years of stability (growth of less than 2.5 mm per year and no worrisome features or high-risk stigmata) was used to establish the reference label for low-risk cases (\textit{N}=117).}}

\end{table}

  \begin{figure}[htbp]
     \centering
     \includegraphics[width=0.8\linewidth]{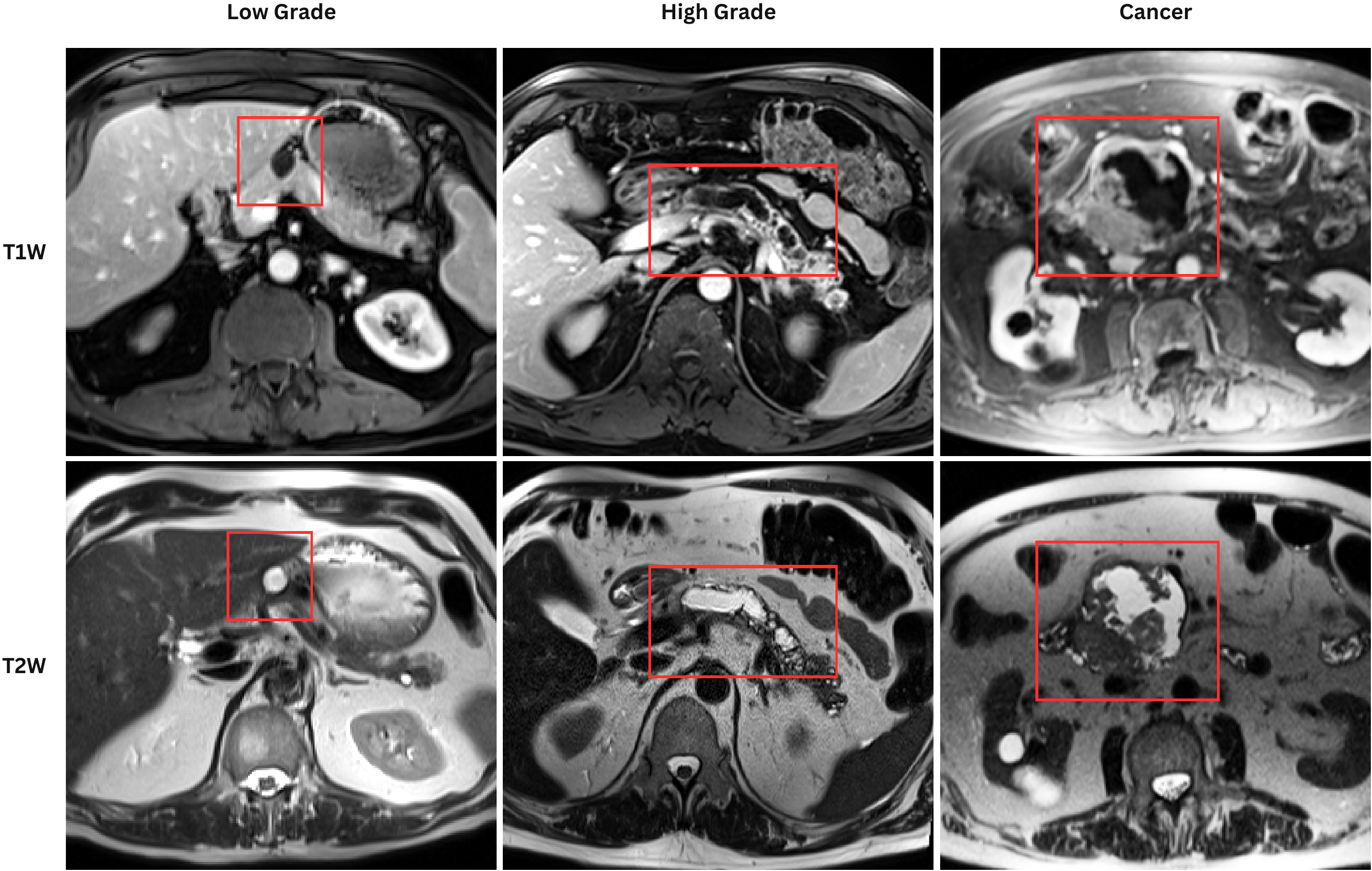}
     \caption{{IPMNs with low-grade dysplasia, high-grade dysplasia, and invasive carcinoma} from the \textit{Cyst-X} dataset.}
     \label{fig: cyst}
 \end{figure}


Fig.~\ref{fig: cyst} shows IPMN examples with low-grade dysplasia, high-grade dysplasia, and invasive carcinoma from the \textit{Cyst-X} dataset. To evaluate variability in image acquisition, we applied uniform manifold approximation and projection (\textit{UMAP}) to image quality indicators, revealing distinct clustering patterns by imaging center and slice thickness {(Fig.~\ref{fig: MRQy}; Section~\ref{sec:Inter-center heterogeneity and its imaging correlates})}. {This heterogeneity reflects real-world clinical variability; it makes the dataset more representative but also poses technical challenges for model development. Because the data were acquired with diverse imaging devices and acquisition protocols, we extracted 21 image-quality indicators with the open-source \textit{MRQy} tool~\cite{sadri2020mrqy}. These included statistical measures of intensity (\textit{e.g.}, mean, range, variance) and second-order or filter-based metrics (\textit{e.g.}, contrast per pixel, entropy focus criterion, and signal-to-noise ratios).}


\subsection{Preprocessing and pancreas segmentation}\label{sec:Preprocessing and pancreas segmentation}
All MRI scans were converted from DICOM to NIfTI format. Expert radiologists at each participating institution manually segmented the pancreas to establish ground-truth annotations using a standardized protocol in ITK-SNAP~\cite{yushkevich2016itk}; a senior radiologist reviewed every mask for quality and consistency. {The annotation protocol included the entire pancreatic region, including cystic lesions, exophytic cyst components, and associated masses. These regions therefore provided lesion-associated information as well as pancreatic parenchyma to the classifiers.} For automated segmentation, we applied \textit{PanSegNet}~\cite{zhang2025large} (Fig.~\ref{fig: overview}), the \textit{nnU-Net}-based~\cite{isensee2021nnu} pancreas segmentation architecture published by our group. {It incorporates linear self-attention~\cite{shen2021efficient} to enhance global context modeling while preserving linear computational complexity, and it was used without architectural or training-objective modification from its original configuration}~\cite{zhang2025large}. 
\textit{PanSegNet} was not federated in the present study because its \textit{nnU-Net} planning procedure uses centralized dataset-level statistics and would require modification or federated fingerprint aggregation. We therefore used \textit{Swin-UNETR}~\cite{hatamizadeh2021swin} for the controlled federated segmentation experiments {(implemented in PyTorch~\cite{paszke2019pytorch} and MONAI~\cite{cardoso2022monai})}. Both segmentation networks were trained and evaluated using five-fold cross-validation{, and \textit{PanSegNet} was additionally evaluated with leave-one-center-out cross-validation}. {Details of \textit{PanSegNet} and linear attention are presented in Supplementary Section~\ref{sup:pansegnet}.}

\subsection{Radiologist visual scoring}\label{sec:Radiologist visual scoring}
{To benchmark the models against expert radiologists on the same patients under matched imaging-only conditions,} three abdominal radiologists independently scored a defined subset of the cohort using the imaging features specified by the Kyoto criteria~\cite{ohtsuka2024international}. {The radiologists had seven, four, and four years of subspecialty experience, and each had interpreted more than 2,000 abdominal MRI examinations.} From the full cohort, 67 patients were excluded for lacking paired T1W and T2W sequences and 68 for lacking ground-truth risk labels, yielding 629 paired cases in total for radiologist evaluation. Radiologists were blinded to histopathology, clinical history, laboratory values and prior imaging, and were instructed to assign each case to one of the three risk categories. {For the binary analyses, the no-risk and low-risk categories were combined as non-high-risk, and the high-risk category was treated as high-risk.} Their predictions were subsequently analyzed in the full 629-case broad cohort, the 474-case IPMN-only subset, {and the 357-case histopathology-confirmed subset}. This protocol simulates the imaging-only initial assessment that radiology workflows produce. {Radiologists reviewed full T1W and T2W volumes, and the models received radiologist-defined pancreas ROIs from T1W, T2W, or both sequences (Table~\ref{tab: ai vs human}). The protocol} 
{does not reproduce the broader clinical decision-making context, including symptoms, serum tumor markers, and multidisciplinary review, that routine guideline application incorporates.}


\subsection{Radiomics and deep learning approaches for IPMN classification}\label{sec: DenseNet-121 for IPMN Classification}
{Models were evaluated on the three classification tasks defined in Section~\ref{sec: results classification}: IPMN-only (primary), broad binary and three-class; Table~\ref{tab: IPMN Classification Results} lists the models evaluated for each task. Class counts per center are given in Supplementary Table~\ref{tab: classification distribution}.} {Center MCA contributed no patients to the no-risk group and center AHN only one, so the no-risk group is unevenly represented across centers.}


The radiomics pipeline characterizes the entire three-dimensional pancreatic {region of interest} using a panel of 1,409 features per scan{, extracted with \textit{PyRadiomics}~\cite{vangriethuysen2017computational}: shape features of the ROI, and first-order intensity, gray-level co-occurrence matrix, gray-level run length matrix, gray-level size zone matrix, gray-level dependence matrix and neighboring gray tone difference matrix features computed on the original image and on wavelet-decomposed, gradient, local binary pattern, square, square-root, logarithm and exponential filtered images.} Prior to extraction, all scans underwent N4 bias-field correction~\cite{tustison2010n4itk}, intensity normalization, and isotropic resampling to 1-mm voxels. Feature selection used maximum-relevance minimum-redundancy (mRMR) filtering~\cite{peng2005feature} applied within each cross-validation fold on the training partition only, with the final feature subset used to train a random-forest classifier~\cite{ho1995random}.

The 3D \textit{DenseNet-121} architecture~\cite{huang2017densely} {(trained from scratch)} was selected as the principal deep-learning classifier for \textit{Cyst-X} because it {had the highest mean AUC among the evaluated} 3D backbones (\textit{ResNet-34}, \textit{ResNet-50}~\cite{he2016deep,hara2018can}, and \textit{EfficientNet-B0}~\cite{tan2019efficientnet}) {for broad binary (high-risk versus non-high-risk) and three-class classification on both T1W and T2W MRI. \textit{DenseNet-121} was therefore used in all subsequent analyses.}
{For the deep-learning classifiers, each scan was intensity-normalized, isotropically resampled, and masked to retain only voxels within the pancreas segmentation.} Z-score standardization was then applied to the masked voxels, and the pancreas-bounded ROI was cropped and resized to $96\times96\times96$ {voxels} (see Supplementary Section \ref{sup:Deep learning models training} for more details).

{We also evaluated several fusion strategies to integrate T1W and T2W information (Supplementary Section~\ref{sup:fusion strategies}).} We compared the predictions of the {radiomics, \textit{DenseNet-121}, and fusion} models with the radiologists' assessments. {For each test case, the prediction came from the cross-validation model in which that case was held out, and the resulting labels were pooled for comparison with the radiologist assessments.}

\subsection{Federated learning implementation}\label{sec:FLimplementation}
{\textit{PanSegNet} was not federated, for the reasons stated in Section~\ref{sec:Preprocessing and pancreas segmentation}. Federated classification experiments used \textit{DenseNet-121} trained on radiologist-defined ROIs. 
Separate federated segmentation experiments used \textit{Swin-UNETR} ({Section~\ref{sec:segmentation results}}). We implemented \textit{FedAvg}~\cite{mcmahan2017communication} and \textit{FedProx}~\cite{li2020federated} and simulated them on a centralized server by partitioning the {multicenter} dataset into seven institutional silos, one per center. Implementation details are presented in Supplementary Section~\ref{sup:fedavg and fedprox}.} 

\subsection{Statistical analysis and visual attribute maps}\label{sec:statistical analysis}
Segmentation performance was reported as Dice coefficient, Jaccard index, precision, recall, 95th-percentile Hausdorff distance (HD95), and average symmetric surface distance (ASSD). Classification performance was reported as area under the curve (AUC) with 95\% Confidence Interval (95\% CI), average precision (the area under the precision-recall curve), accuracy (ACC), sensitivity (Sens), and specificity (Spec). {Pooled AUC was computed by combining test predictions across all institutions. For external leave-one-center-out evaluation, unweighted center-averaged AUCs (the mean of the seven center-specific AUCs) are also reported to assess inter-institutional generalizability. AUC 95\% CIs were calculated using percentile bootstrapping (1,000 iterations). Cases were resampled with replacement within each evaluation cohort or test fold; for cross-validation, fold-specific AUCs were averaged without weighting. Bootstrap replicates lacking any class present in the original sample were excluded. Confidence limits were the 2.5th and 97.5th percentiles of the retained estimates. For external comparisons of federated and centralized models, 95\% CIs for the paired differences in pooled and center-averaged AUC were obtained from 1,000 bootstrap replicates in which patients were resampled within each center and true class, using the same resampled patients for all models.} {To assess whether discrimination in the IPMN-only task depended on between-center differences in prevalence, we fitted logistic regression models with center as a fixed effect and the standardized model score as a covariate; model scores were rank-normalized (within each cross-validation fold for internal evaluation and across all held-out predictions for external evaluation) before standardization, and odds ratios are reported per standard deviation of the score with Wald 95\% CIs.}

{Calibration of pooled test predictions was assessed by the calibration slope (the coefficient of a logistic regression of the outcome on the logit of the predicted probability; ideal value 1), the calibration-in-the-large intercept (estimated with the logit as an offset; ideal value 0) and the Brier score. Because each test case was scored by the fold-specific or center-specific model in which it was held out, each model was also recalibrated by logistic regression on the logit of its own validation predictions, and the fitted transformation was applied to that model's test predictions; test labels and data from the held-out center were not used. Clinical utility was assessed by decision-curve analysis~\cite{vickers2006decision}, comparing the net benefit of the recalibrated models with treating all or no patients at threshold probabilities of 0.10--0.40. 95\% CIs for calibration measures and for differences in net benefit were obtained from 1,000 patient-level bootstrap replicates stratified by true class.}
{Comparisons across architectures, modalities, segmentation inputs, fusion strategies and operating thresholds were not adjusted for multiplicity and are reported descriptively.}


{To visualize image regions associated with classifier predictions,} we generated \textit{Grad-CAM}~\cite{selvaraju2017grad} and \textit{IBA}~\cite{schulz2020restricting,demir2021information} maps for representative test cases. \textit{Grad-CAM} highlights spatial regions that strongly influence the model’s decision by computing gradients of the target class score with respect to feature activations. In contrast, \textit{IBA} {learns} a perturbation mask {that restricts the information passing through an intermediate layer while preserving the information needed for the prediction}, {thereby highlighting features associated with the prediction. In the illustrated cases, \textit{IBA} maps were more spatially concentrated than \textit{Grad-CAM} maps (Supplementary Fig.~\ref{fig: iba}).} 


\section{Data Availability}\label{sec:data_availability_split}
The \textit{Cyst-X} dataset, including anonymized MRI scans, segmentation masks, risk labels, {and fold splits}, is publicly available at {Hugging Face: \url{https://huggingface.co/datasets/phy710/Cyst-X/}. We also provide the mirror on OSF: \url{https://osf.io/74vfs/}. The analyses reported here used \textit{Cyst-X} dataset revision v2), with patient-level five-fold internal classification splits and leave-one-center-out evaluation. The corresponding split files are \texttt{classification/t1\_split.csv} and \texttt{classification/t2\_split.csv}, with paired T1W--T2W fusion analyses using the T1W assignments; earlier four-fold binary classification splits belong to the v1 release and were not used for these analyses.}

\section{Code Availability}
Source code for the \textit{PanSegNet} segmentation model and classification algorithms is available at \url{https://github.com/NUBagciLab/Cyst-X}. {Model weights and segmentation results are available at \url{https://huggingface.co/phy710/Cyst-X}. The reported results correspond to code release v2, dataset revision v2, and model revision v2. A release manifest at \url{https://github.com/NUBagciLab/Cyst-X/blob/v2/README.md} identifies the preprocessing configuration, task-specific split files, model checkpoints, and evaluation commands used to generate the reported tables and figures.}

\section{Acknowledgments}
This work was supported by NIH NCI R01-CA246704, R01-CA240639, U01-CA268808, and NIH-NHLBI R01-HL171376.

\section{Author Contributions}

U.B., H.P., G.D. and H.E.A. designed the experiments; H.P. trained the Swin-UNETR for segmentation, implemented federated learning for segmentation and classification. {Z.H. and Z.Z. developed the \textit{PanSegNet} model for pancreas segmentation. Z.H. trained radiomics-based IPMN classification.} {Z.W. performed the additional statistical analyses and contributed to manuscript revision.} G.D. and E.K. coordinated multi-institutional data collection and contributed to manuscript preparation. G.D., E.K., H.E.A., and A.M.B. organized and curated the datasets. D.S., A.M., Y.T., G.D.K., M.S.E., T.C., E.A., P.T., S.J., I.G.S., M.J.B., C.H., C.W.B., T.G., F.H.M., R.N.K., and M.B.W. contributed to data collection and clinical validation. Y.V., L.Z., Z.X., F.P.S., and C.S. advised on models' implementation, evaluation, and statistical analysis. M.B.W. and U.B. supervised the project, guided the experimental design, and wrote and critically revised the manuscript. Radiologist studies were guided by G.D. and H.E.A. All authors discussed the results and approved the final version.

\section{Competing Interests}
The authors declare no competing interests.



\bibliography{sn-bibliography}

\newpage
\begin{appendices}
\section{Supplementary Materials}

\subsection{{Uniform manifold approximation and projection}}\label{sup:UMAP}
{To calculate the \textit{UMAP} for dimension reduction, each feature was normalized across the dataset using the following techniques:
\begin{itemize}
    \item MinMax normalization:
    \begin{equation}
    \label{eq:minmax}
    x_{minmax}' = \frac{x-x_{min}}{x_{max}-x_{min}},    
\end{equation}
\item Variance normalization: 
    \begin{equation}
    \label{eq:white}
    x_{variance}' = \frac{x}{\sigma},    
\end{equation}
\item Z-score normalization:
    \begin{equation}
    \label{eq:zscore}
    x_{zscore}' = \frac{x-m}{\sigma},
\end{equation}
\end{itemize}
where $x$ and  $x'$ are the original and normalized values, respectively, while {$m$} and $\sigma$ are the mean and standard deviation of the dataset.}

\subsection{{\textit{PanSegNet} and linear attention}}\label{sup:pansegnet}
{Let $\mathbf{X}\in\mathbb{R}^{N\times d}$ denote the input feature matrix, where each row corresponds to an individual feature vector. Let $\mathbf{A}, \mathbf{A'}\in\mathbb{R}^{N\times d}$ represent the outputs of traditional and linear self-attention mechanisms, respectively. Here, $N$ and $d$ represent the length and the dimension of the feature, with $N>d$.
The queries, keys, values $\mathbf{Q, K, V}\in\mathbb{R}^{N\times d}$ are computed by linearly projecting the input features $\mathbf{X}$:
\begin{align}
\mathbf{Q} &= \mathbf{X}\mathbf{W}_Q + \mathbf{b}_Q, \\
\mathbf{K} &= \mathbf{X}\mathbf{W}_K + \mathbf{b}_K, \\
\mathbf{V} &= \mathbf{X}\mathbf{W}_V + \mathbf{b}_V,
\end{align}
where $\mathbf{W}_Q$, $\mathbf{W}_K$, $\mathbf{W}_V\in\mathbb{R}^{d\times d}$ and $\mathbf{b}_Q$, $\mathbf{b}_K$, $\mathbf{b}_V\in\mathbb{R}^d$ are learnable parameters.
The traditional self-attention is defined as:
\begin{equation}
    \mathbf{A}_i = \sum_{j=0}^{N-1}sim(\mathbf{Q}_i, \mathbf{K_j})\mathbf{V}_j,
\end{equation}
where $i, j$ index the input features, and the similarity function is defined as:
\begin{equation}
sim(\mathbf{q}, \mathbf{k})=SoftMax\left(\frac{\mathbf{qk}^T}{\sqrt{d}}\right).
\end{equation}
In linear self-attention~\cite{shen2021efficient}, the similarity function is approximated by decomposing it into two kernel feature maps: \begin{equation}
sim'(\mathbf{q}, \mathbf{k}) = \phi(\mathbf{q})\rho(\mathbf{k})^T.\end{equation}
Here, $\phi(\cdot)$ and $\rho(\cdot)$ apply the SoftMax function to the queries row-wise and keys column-wise, respectively. 
The resulting linear self-attention is then computed as:
\begin{equation}
    \mathbf{A}'_i = \sum_{j=0}^{N-1}\phi(\mathbf{Q}_i)\rho(\mathbf{K}_j)^T\mathbf{V}_j = \phi(\mathbf{Q}_i)\sum_{j=0}^{N-1}\rho(\mathbf{K}_j)^T\mathbf{V}_j.
\end{equation}
This definition keeps the important property of original self-attention, \textit{i.e.}, $\sum_{j=0}^{N-1}sim(\mathbf{Q}_i, \mathbf{K_j})=1$ and reduces the computational complexity from $O(dN^2)$ to $O(d^2N)$.}

{Training of \textit{PanSegNet} followed the implementation described in~\cite{zhang2025large}. {We used stochastic gradient descent (\textit{SGD}) with a learning rate of 0.01, a batch size of 2, and 600 training epochs.} To mitigate overfitting, we incorporated dropout layers and applied data augmentation techniques during training.}

\subsection{Deep learning model training}\label{sup:Deep learning models training}
We employed the \textit{AdamW} optimizer~\cite{loshchilov2017decoupled} with an initial learning rate of 0.001 and a batch size of 16 {for a maximum of 100 epochs}. The learning rate was progressively reduced by a factor of 10 every 30 epochs to enhance convergence. {Internal evaluation used five-fold cross-validation with fixed patient-level fold assignments stratified jointly by center and three-class risk label, so that the patients of each center and risk class were divided as evenly as possible across the five folds (Section~\ref{sec:data_availability_split}); one fold was held out for testing in each iteration. The IPMN-only analyses used the same assignments after excluding no-risk patients, and the histopathology-confirmed analyses after additionally excluding surveillance-defined patients. External evaluation used leave-one-center-out cross-validation, pooling data from six centers for model development and holding out the remaining center for testing, with each center serving once as the test center. In both settings, the development data were divided at the patient level into training and validation subsets in a 4:1 ratio, using a class-stratified random split within each center (random seed 42). The validation subset was used for early stopping and decision-threshold selection, while the test subset was reserved for performance evaluation. Patients did not overlap among the training, validation, and test subsets.}


\subsection{Fusion strategies}\label{sup:fusion strategies}
{Suppose $\mathbf{x}_1, \mathbf{x}_2 \in \mathbb{R}^{1 \times H \times W \times D}$ represent T1W and T2W volumetric inputs from the same patient, respectively, and $y \in [0, 1]$ denotes the predicted probability of an IPMN high-risk lesion. Let $f(\cdot)$ denote a 3D \textit{DenseNet-121} feature extractor mapping volumetric inputs to latent embedding vectors, $h(\cdot)$ denote a linear classification head outputting a scalar logit, and $\sigma(\cdot)$ denote the sigmoid activation function. We evaluate four multimodal fusion paradigms:
\begin{itemize}
    \item \textbf{Early Fusion:} Channel-wise concatenation at the input layer, processed by a single network with a two-channel first convolutional layer:
    \begin{equation}
        y = \sigma\left(h\left(f(\mathbf{x}_1 \mathbin{\Vert} \mathbf{x}_2)\right)\right),
    \end{equation}
    where $\mathbin{\Vert}$ denotes concatenation along the channel dimension.
    \item \textbf{Late Fusion:} Modality-specific encoders $f_1(\cdot)$ and $f_2(\cdot)$ with separate parameters, {trained jointly with} a shared classification head {applied to their concatenated latent representation vectors}:
    \begin{equation}
        y = \sigma\left(h\left(f_1(\mathbf{x}_1) \mathbin{\Vert} f_2(\mathbf{x}_2)\right)\right).
    \end{equation}
    \item \textbf{Siamese Fusion:} A shared-weight encoder $f(\cdot)$ extracting features independently from each modality, followed by latent-vector concatenation and a joint classification head:
    \begin{equation}
        y = \sigma\left(h\left(f(\mathbf{x}_1) \mathbin{\Vert} f(\mathbf{x}_2)\right)\right).
    \end{equation}
    \item \textbf{Logit Fusion:} Two complete, independent single-modality networks $(f_1, h_1)$ and $(f_2, h_2)$ whose unnormalized scalar logits are blended via a learnable weighting coefficient $\alpha \in [0, 1]${; $\alpha$ and $1-\alpha$ are obtained by applying a softmax to two learnable scalars}:
    \begin{equation}
        y = \sigma\left(\alpha h_1(f_1(\mathbf{x}_1)) + (1 - \alpha) h_2(f_2(\mathbf{x}_2))\right).
    \end{equation}
\end{itemize}}

\subsection{\textit{FedAvg} and 
\textit{FedProx}}\label{sup:fedavg and fedprox}
{Our earlier studies explored federated learning for IPMN risk assessment and pancreas MRI segmentation~\cite{pan2025ipmn,pan2025adaptive}. All experiments reported in the present study were rerun using the current Cyst-X cohort and evaluation protocols.}
{\textit{FedAvg} aggregates model updates by averaging weights from participating clients. Formally, the update rule for the global model $\mathbf{w}^t$ at {communication round} $t$ is: 
\begin{equation}
        \mathbf{w}^t = \frac{\sum_{k=0}^{K-1} N_k \mathbf{w}_k^t}{\sum_{k=0}^{K-1} N_k},
    \end{equation}
where $\mathbf{w}_k^t$ is the model {trained by the $k$-th client on its own data}, $N_k$ is the number of samples from the $k$-th client, and $K$ is the number of participating clients.
In contrast, \textit{FedProx} introduces a proximal term $\frac{\mu}{2} \| \mathbf{w}^{t+1}_k - \mathbf{w}^t \|^2$ to the loss function of each client $\ell_k$ to mitigate the impact of client drift and heterogeneous data, allowing for more stable convergence:
    \begin{equation}
        \ell_k^{\text{prox}}(t+1) = \ell_k(t+1) + \frac{\mu}{2} \| \mathbf{w}_k^{t+1} - \mathbf{w}^t \|^2.
    \end{equation}
{We evaluated several values of the proximal term $\mu$.} {For both algorithms, each communication round consisted of one local training epoch followed by aggregation, for a total of 100 rounds.}} 

\subsection{{Validation-based operating-threshold selection}}\label{sec:Clinically meaningful threshold calibration}
{Decision thresholds were selected using validation data to maximize sensitivity subject to a specificity target of at least 90\%. This target was chosen after observing the specificities of the three blinded radiologists on the broad 629-case cohort (89.94\%, 97.13\%, and 94.66\%; mean, 93.91\%; Table~\ref{tab: ai vs human}); 90\% approximates the lower end of this range.
Candidate thresholds consisted of the distinct predicted validation probabilities. A case was classified as high-risk when its predicted probability was greater than or equal to the threshold. We selected the lowest threshold achieving validation specificity of at least 90\%, which yields the highest sensitivity among the candidate thresholds satisfying this constraint. 
A separate threshold was selected for each trained model and evaluation split using its validation subset, rather than by tuning thresholds separately on each test center. For internal cross-validation, threshold selection used the validation subset drawn from the four development folds. For external leave-one-center-out evaluation, it used the validation subset drawn from the six development centers. Numerical thresholds were computed from validation data and fixed before evaluation on the corresponding test data; neither test labels nor data from the held-out center entered this computation. Results obtained with these validation-selected thresholds are reported alongside results at the fixed 0.5 threshold in Table~\ref{tab: ai vs human} and Supplementary Tables~\ref{tab: t1 low vs high classification}--\ref{tab: fusion full}.}

{Operating-threshold selection on validation data does not guarantee the same operating characteristics on test data, particularly when center-level sample sizes are small or cohort distributions differ. For example, the external T2W \textit{DenseNet-121} results at NU showed 85.56\% accuracy and 92.40\% specificity but only 12.50\% sensitivity (Supplementary Table~\ref{tab: t2 classification}). High accuracy in such an imbalanced cohort should therefore be interpreted alongside sensitivity and specificity. Alternative threshold-selection objectives, including net-benefit optimization~\cite{vickers2006decision} and cost-sensitive thresholding~\cite{elkan2001foundations}, warrant future investigation.}

\subsection{{Additional analyses of radiologist comparisons and classification performance}}
\label{sec:zw-paired-analysis}
{We used the saved held-out predictions for three additional analyses: paired comparisons with radiologists, evaluation on a common histopathology-confirmed subset, and internal--external AUC comparisons using consistent aggregation methods. The analyses used the original fitted models and validation-selected thresholds.}

\paragraph{{Paired comparisons with radiologists.}}
{We matched model predictions to radiologist assessments by patient identifier. For the T2W DenseNet-121 models trained separately for each task and evaluated by internal cross-validation, we evaluated the broad paired cohort ($n=629$), IPMN-only cohort ($n=474$), and histopathology-confirmed IPMN subset ($n=357$). We calculated sensitivity and specificity for each model and the arithmetic mean of the corresponding metrics across the three radiologists (Supplementary Table~\ref{tab:zw-reader-ci}). Paired differences were calculated as the model metric minus the radiologist mean, in percentage points (Supplementary Table~\ref{tab:zw-paired-differences}). Percentile 95\% confidence intervals were obtained from 1,000 patient-level bootstrap replicates stratified by true risk class. Within each replicate, the same sampled patients were used for the model and all three radiologists, and the radiologist mean was recomputed. Bootstrap intervals summarize patient-sampling uncertainty for the fixed models, thresholds, centers, and radiologists. Intervals are unadjusted for multiple comparisons.}

\paragraph{{Evaluation on a common histopathology-confirmed subset.}}
{We restricted the saved broad-task and IPMN-only predictions to the same 357 histopathology-confirmed patients (142 high-risk and 215 low-risk), retaining the original models and validation-selected thresholds. We also evaluated the saved predictions from the separately trained histopathology-only models on these patients (Supplementary Table~\ref{tab:zw-frozen-pathology}). This provides a common evaluation population while distinguishing the original models from models trained specifically on histopathology-confirmed cases. Sensitivity and specificity use the retained thresholds; AUCs use pooled held-out probabilities. For all rows, percentile 95\% AUC confidence intervals were calculated from 1,000 patient-level bootstrap replicates stratified by true risk class, with fitted models held fixed. Tied probabilities receive half credit in pairwise AUC comparisons.}

\paragraph{{Internal--external AUC comparisons.}}
{For the broad binary task, we compared centralized and federated DenseNet-121 models on the same patients within each modality (652 T1W and 655 T2W). Internal predictions were obtained by cross-validation and external predictions by leave-one-center-out evaluation. We calculated pooled AUC across all patients and the unweighted mean of the seven center-specific AUCs, applying each aggregation method consistently in both settings (Supplementary Table~\ref{tab:zw-matched-auc}). Centralized pooled AUC was 74.94\% internally and 71.35\% externally on T1W, and 78.17\% and 77.49\%, respectively, on T2W. The internal pooled AUCs differ from the fold-mean AUCs in Table~\ref{tab: IPMN Classification Results}. Pooled AUC summarizes ranking across patients from all centers, whereas center-averaged AUC summarizes discrimination within centers. Each center-specific AUC was computed once on that center's predictions pooled across folds (internal) or taken from its held-out evaluation (external), and the seven values were then averaged without weighting. In the internal setting, a center's patients are therefore scored by different cross-validation models, so differences in score scale between fold models can lower the center-specific AUC.}

\clearpage
\subsection{Supplementary figures and tables}
\begin{figure}[htbp]
\centering
\includegraphics[width=\linewidth]{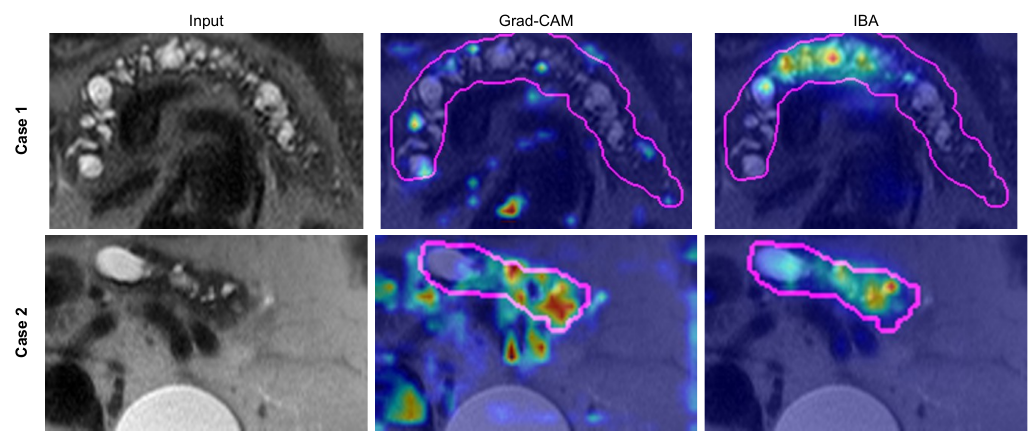}
\caption{Visual explanations using \textit{Grad-CAM} and \textit{IBA} {for two representative cases.}}
\label{fig: iba}
\end{figure}

\begin{figure}[htbp]
\centering
\includegraphics[width=\linewidth]{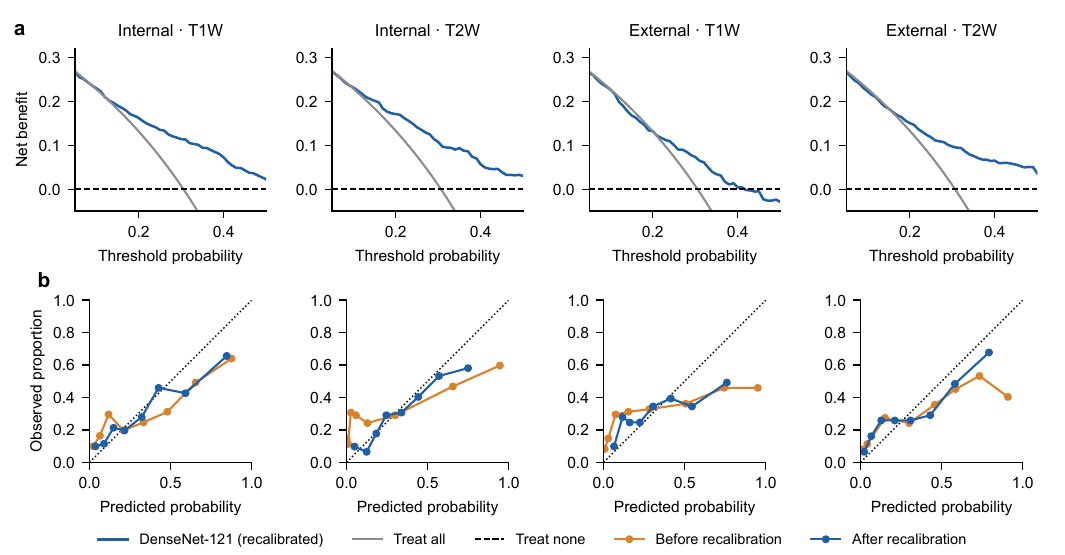}
\caption{{Decision-curve and calibration analysis of \textit{DenseNet-121} for the primary IPMN-only task under internal cross-validation and external leave-one-center-out evaluation. (a) Net benefit of the recalibrated models compared with treating all or no patients. (b) Calibration curves before and after validation-based recalibration; points show the mean predicted probability and observed proportion of high-risk IPMN in eight quantile bins, and the dotted line indicates perfect calibration. Each fold-specific or held-out-center-specific model was recalibrated on its own validation predictions only.}}
\label{fig: calibration}
\end{figure}

\begin{table}[htbp]
\caption{Data distribution for IPMN classification. {No, Low and High denote no-risk, low-risk IPMN and high-risk IPMN.}}
\centering
\aboverulesep=0ex
\belowrulesep=0ex
\setlength{\tabcolsep}{3pt}
{
\begin{tabular}{l|llll|llll|llll}
\toprule
\belowrulesepcolor{g1}
\rowcolor{g1}&\multicolumn{4}{l|}{\textbf{T1W Modality}}&\multicolumn{4}{l|}{\textbf{T2W Modality}}&\multicolumn{4}{l}{\textbf{Paired Two Modalities}}\\
\rowcolor{g1}\multirow{-2}{*}{\textbf{Data Centers}}&No&Low& High&Total&No&Low& High&Total&No&Low& High&Total\\
\hline
\rowcolor{g2}Center 1: NYU& 48 & 79 & 23 & 150 & 48 & 79 & 24 & 151 & 48 & 79 & 23 & 150\\
\rowcolor{g2}Center 2: MCF& 29 & 42 & 63 & 134 & 25 & 42 & 63 & 130 & 25 & 42 & 63 & 130\\
\rowcolor{g2}Center 3: NU& 43 & 126 & 17 & 186 & 44 & 127 & 16 & 187 & 43 & 126 & 16 & 185\\
\rowcolor{g2}Center 4: AHN& 1 & 11 & 4 & 16 & 1 & 13 & 4 & 18 & 1 & {11} & {3} & 15\\
\rowcolor{g2}Center 5: MCA& 0 & 10 & 14 & 24 & 0 & 7 & 16 & 23 & 0 & 6 & 9 & 15\\
\rowcolor{g2}Center 6: IU & 3 & 48 & 13 & 64 & 3 & 46 & 14 & 63 & 3 & 46 & 13 & 62\\
\rowcolor{g2}Center 7: EMC& 40 & 23 & 15 & 78 & 38 & 30 & 15 & 83 & 35 & 22 & 15 & 72\\
\hline
\rowcolor{g3}\textbf{Total} & 164 & 339 & 149 & 652 & 159 & 344 & 152 & 655 & 155 & {332} & {142} & 629\\
\aboverulesepcolor{g3}
\bottomrule
\end{tabular}}
\label{tab: classification distribution}
\end{table}

\begin{table}[htbp]
    \centering
    \caption{{\textit{PanSegNet}'s internal evaluation performance. {Results are shown as mean $\pm$ s.d. across patients pooled over the five cross-validation folds.}}}
    \begin{tabular}{lllllll}
    \toprule
    \belowrulesepcolor{g1}
    \rowcolor{g1}\bf{Center}&\bf{Dice(\%)} & \bf{Jaccard(\%)} & \bf{Precision(\%)} &\bf{Recall(\%)} & \bf{HD95(mm)} & \bf{ASSD(mm)} \\\hline
\rowcolor{g3}\multicolumn{7}{l}{T1W Modality}\\\hline
\rowcolor{g2}NYU &   84.99$\pm$7.74 &   74.56$\pm$9.84 &   84.80$\pm$8.66 &   86.10$\pm$8.32 & {2.93$\pm$5.36} & {0.63$\pm$1.96} \\
\rowcolor{g2}MCF & {87.48$\pm$5.79} & {78.17$\pm$8.36} & {87.64$\pm$7.44} &   88.11$\pm$7.37 & {2.98$\pm$5.78} & {0.51$\pm$0.48} \\
\rowcolor{g2}NU &   88.33$\pm$5.89 &   79.52$\pm$7.98 &   87.49$\pm$7.69 &   89.72$\pm$5.74 & {2.94$\pm$6.22} & {0.64$\pm$0.82} \\
\rowcolor{g2}AHN &   81.22$\pm$6.11 &   68.79$\pm$8.16 &   84.54$\pm$8.92 &   79.15$\pm$7.85 & {4.25$\pm$3.85} & {0.56$\pm$0.52} \\
\rowcolor{g2}MCA &  78.72$\pm$10.98 &  66.03$\pm$12.55 &  75.29$\pm$14.90 &   85.04$\pm$6.52 & {7.44$\pm$12.06} & {1.87$\pm$4.48} \\
\rowcolor{g2}IU &   85.37$\pm$9.91 &  75.56$\pm$12.65 &   86.13$\pm$9.51 &  85.54$\pm$11.58 & {5.61$\pm$9.14} & {1.05$\pm$2.14} \\
\bottomrule
    \end{tabular}
    \label{tab: panseg}
\end{table}

\begin{table}[htbp]
    \centering
    \caption{{\textit{PanSegNet}'s external evaluation performance.}}
    {
    \begin{tabular}{lllllll}
    \toprule
    \belowrulesepcolor{g1}
    \rowcolor{g1}\bf{Center}&\bf{Dice(\%)} & \bf{Jaccard(\%)} & \bf{Precision(\%)} &\bf{Recall(\%)} & \bf{HD95(mm)} & \bf{ASSD(mm)} \\\hline
\rowcolor{g3}\multicolumn{7}{l}{T1W Modality}\\\hline
\rowcolor{g2}NYU &   84.62 &   73.88&   85.66&   84.41 & {3.20} & {0.50} \\
\rowcolor{g2}MCF &   86.93 &   77.38 &   87.42 &   87.32 & {3.19} & {0.51} \\
\rowcolor{g2}NU &   87.98&   79.00 &   86.54&   90.07 & {2.82} & {0.69} \\
\rowcolor{g2}AHN &   80.55&   67.85 &   85.32 &   77.44 & {3.97} & {0.54} \\
\rowcolor{g2}MCA &   77.66&   64.70 &   73.58 &   85.47 & {8.16} & {2.10} \\
\rowcolor{g2}IU &   85.00 &   75.11 &   85.24 &   85.56 & {5.34} & {1.07} \\
\rowcolor{g2}EMC &   89.28 &   80.99 &   89.34&   89.76 & {4.82} & {0.66} \\
\hline
\rowcolor{g2}Global &   86.34 &  76.62 &   86.27&   87.28& {3.70} & {0.69} \\
\aboverulesepcolor{g2}\midrule\belowrulesepcolor{g3}
\rowcolor{g3}\multicolumn{7}{l}{T2W Modality}\\\hline
\rowcolor{g2}NYU &   86.76 &   76.90 &   88.19 &   85.97 & {2.93} & {0.45} \\
\rowcolor{g2}MCF &   87.43 &   78.12&   88.78 &   87.02 & {2.97} & {0.42} \\
\rowcolor{g2}NU &   92.01 &   85.63 &   90.73&   93.82 & {2.01} & {0.31} \\
\rowcolor{g2}AHN &   82.36 &   71.31 &   85.65 &   80.51 & {4.17} & {1.03} \\
\rowcolor{g2}MCA &   84.53 &   73.38 &   88.69 &   81.64 & {3.25} & {0.38} \\
\rowcolor{g2}IU &   92.29 &   85.91 &   94.30 &   90.55& {1.56} & {0.19} \\
\rowcolor{g2}EMC &   90.95 &   84.21 &   92.75 &   90.03 & {4.64} & {0.42} \\
\hline
\rowcolor{g2}Global &   89.26 & 81.18  &   90.18 &   89.06 & {2.83} & {0.39} \\
\aboverulesepcolor{g3}
    \bottomrule
    \end{tabular}}
    \label{tab:panseg external}
\end{table}

\begin{table}[htbp]
    \centering
    \caption{\textit{Swin-UNETR}'s performance. {{Results are shown as mean $\pm$ s.d. of the five fold-level means.}}}
    \begin{tabular}{lllllll}
    \toprule
    \belowrulesepcolor{g1}
    \rowcolor{g1}\bf{Center}&\bf{Dice(\%)} & \bf{Jaccard(\%)} & \bf{Precision(\%)} &\bf{Recall(\%)} & \bf{HD95(mm)} & \bf{ASSD(mm)} \\\hline
\rowcolor{g3}\multicolumn{7}{l}{T1W Modality}\\\hline
\rowcolor{g2}NYU&77.78$\pm$2.40 & 65.26$\pm$2.51 & 77.48$\pm$3.78 & 80.72$\pm$1.68 & 23.76$\pm$11.55 & 6.93$\pm$5.58\\
\rowcolor{g2}MCF&81.63$\pm$2.92 & 70.17$\pm$3.51 & 82.97$\pm$2.06 & 82.05$\pm$3.48 & 11.27$\pm$2.84 & 2.68$\pm$1.71\\
\rowcolor{g2}NU&82.89$\pm$2.06 & 72.00$\pm$2.89 & 82.21$\pm$2.55 & 84.90$\pm$1.77 & 13.77$\pm$4.98 & 3.34$\pm$0.96\\
\rowcolor{g2}AHN &65.14$\pm$11.05 & 50.90$\pm$10.48 & 74.43$\pm$6.51 & 61.04$\pm$12.47 & 22.51$\pm$15.68 & 4.23$\pm$2.94\\
\rowcolor{g2}MCA&67.50$\pm$6.05 & 52.45$\pm$6.56 & 66.26$\pm$4.72 & 70.99$\pm$7.94 & 63.22$\pm$28.78 & 13.67$\pm$6.59\\
\rowcolor{g2}IU&79.17$\pm$2.31 & 67.09$\pm$3.06 & 78.74$\pm$3.73 & 81.87$\pm$3.43 & 30.53$\pm$6.75 & 8.03$\pm$2.95\\
\bottomrule
    \end{tabular}
    \label{tab: SwinUnetR}
\end{table}

\begin{table}[htbp]
    \centering
    \caption{\textit{Swin-UNETR}+\textit{FedAvg}'s performance. {{Results are shown as mean $\pm$ s.d. of the five fold-level means.}}}
    \begin{tabular}{lllllll}
    \toprule
    \belowrulesepcolor{g1}
    \rowcolor{g1}\bf{Center}&\bf{Dice(\%)} & \bf{Jaccard(\%)} & \bf{Precision(\%)} &\bf{Recall(\%)} & \bf{HD95(mm)} & \bf{ASSD(mm)} \\\hline
\rowcolor{g3}\multicolumn{7}{l}{T1W Modality}\\\hline
\rowcolor{g2}NYU&71.84$\pm$2.51 & 58.25$\pm$2.21 & 71.06$\pm$3.45 & 76.67$\pm$1.26 & 34.81$\pm$8.00 & 12.12$\pm$6.18\\
\rowcolor{g2}MCF&77.94$\pm$3.36 & 65.40$\pm$3.85 & 79.16$\pm$3.10 & 79.07$\pm$3.57 & 15.64$\pm$3.96 & 3.60$\pm$1.74\\
\rowcolor{g2}NU&77.61$\pm$2.78 & 65.25$\pm$3.38 & 74.06$\pm$2.82 & 83.66$\pm$1.95 & 31.89$\pm$5.17 & 8.36$\pm$1.83\\
\rowcolor{g2}AHN &51.07$\pm$10.96 & 37.61$\pm$8.85 & 61.60$\pm$4.80 & 48.41$\pm$13.08 & 34.78$\pm$12.52 & 12.68$\pm$8.57\\
\rowcolor{g2}MCA&46.13$\pm$11.55 & 32.28$\pm$10.35 & 40.38$\pm$10.29 & 59.55$\pm$12.69 & 110.16$\pm$36.15 & 44.79$\pm$18.87\\
\rowcolor{g2}IU&71.09$\pm$3.26 & 58.03$\pm$4.01 & 67.60$\pm$3.86 & 80.11$\pm$4.29 & 53.95$\pm$6.37 & 15.16$\pm$3.77\\
\bottomrule
    \end{tabular}
    \label{tab: SwinUnetRFedAvg}
\end{table}

\begin{table}[htbp]
    \centering
    \caption{\textit{Swin-UNETR}+\textit{FedProx}($\mu=0.3)$'s performance. {{Results are shown as mean $\pm$ s.d. of the five fold-level means.}}}
    \begin{tabular}{lllllll}
    \toprule
    \belowrulesepcolor{g1}
\rowcolor{g1}\bf{Center}&\bf{Dice(\%)} & \bf{Jaccard(\%)} & \bf{Precision(\%)} &\bf{Recall(\%)} & \bf{HD95(mm)} & \bf{ASSD(mm)} \\\hline
\rowcolor{g3}\multicolumn{7}{l}{T1W Modality}\\\hline
\rowcolor{g2}NYU&60.31$\pm$2.27 & 46.02$\pm$2.12 & 59.92$\pm$2.67 & 66.25$\pm$0.97 & 81.99$\pm$9.69 & 26.72$\pm$5.39\\
\rowcolor{g2}MCF&71.63$\pm$4.18 & 57.63$\pm$4.54 & 73.65$\pm$4.55 & 72.56$\pm$3.43 & 30.03$\pm$11.76 & 6.65$\pm$3.25\\
\rowcolor{g2}NU&65.07$\pm$3.77 & 51.06$\pm$3.74 & 62.81$\pm$4.73 & 70.57$\pm$2.54 & 91.69$\pm$8.79 & 24.60$\pm$3.51\\
\rowcolor{g2}AHN &36.24$\pm$10.71 & 24.91$\pm$8.01 & 37.94$\pm$9.71 & 38.25$\pm$13.66 & 101.79$\pm$4.39 & 35.16$\pm$10.74\\
\rowcolor{g2}MCA&26.91$\pm$15.07 & 18.28$\pm$11.69 & 23.06$\pm$13.27 & 41.17$\pm$16.93 & 141.22$\pm$30.07 & 74.43$\pm$24.62\\
\rowcolor{g2}IU&49.87$\pm$5.22 & 36.51$\pm$4.64 & 43.31$\pm$5.09 & 64.78$\pm$6.79 & 123.78$\pm$8.26 & 47.66$\pm$5.68\\
\bottomrule
    \end{tabular}

    \label{tab: SwinUnetRFedProx03}
\end{table}

\begin{table}[htbp]
    \centering
    \caption{\textit{Swin-UNETR}+\textit{FedProx}($\mu=0.01)$'s performance. {{Results are shown as mean $\pm$ s.d. of the five fold-level means.}}}
    \begin{tabular}{lllllll}
    \toprule
    \belowrulesepcolor{g1}
\rowcolor{g1}\bf{Center}&\bf{Dice(\%)} & \bf{Jaccard(\%)} & \bf{Precision(\%)} &\bf{Recall(\%)} & \bf{HD95(mm)} & \bf{ASSD(mm)} \\\hline
\rowcolor{g3}\multicolumn{7}{l}{T1W Modality}\\\hline
\rowcolor{g2}NYU&70.16$\pm$2.67 & 56.32$\pm$2.50 & 68.93$\pm$2.96 & 75.42$\pm$1.94 & 48.86$\pm$6.08 & 15.39$\pm$5.71\\
\rowcolor{g2}MCF&77.56$\pm$3.35 & 64.78$\pm$3.86 & 78.55$\pm$2.42 & 78.98$\pm$4.17 & 16.27$\pm$5.43 & 3.90$\pm$1.93\\
\rowcolor{g2}NU&76.02$\pm$3.38 & 63.70$\pm$4.09 & 72.70$\pm$4.04 & 82.47$\pm$2.04 & 38.92$\pm$12.04 & 10.19$\pm$2.66\\
\rowcolor{g2}AHN&47.10$\pm$10.95 & 33.87$\pm$8.76 & 57.15$\pm$7.38 & 44.02$\pm$13.92 & 48.54$\pm$16.35 & 17.19$\pm$9.42\\
\rowcolor{g2}MCA&38.75$\pm$13.79 & 26.78$\pm$11.58 & 33.39$\pm$12.75 & 52.80$\pm$14.80 & 118.66$\pm$40.33 & 58.79$\pm$22.81\\
\rowcolor{g2}IU&67.67$\pm$3.76 & 54.48$\pm$4.85 & 64.12$\pm$4.80 & 77.69$\pm$4.72 & 70.14$\pm$9.79 & 21.67$\pm$4.17\\
\bottomrule
    \end{tabular}
    \label{tab: SwinUnetRFedProx001}
\end{table}

\begin{table}[htbp]
    \centering
    \caption{\textit{Swin-UNETR}+\textit{FedProx}($\mu=0.005)$'s performance. {{Results are shown as mean $\pm$ s.d. of the five fold-level means.}}}
    \begin{tabular}{lllllll}
    \toprule
    \belowrulesepcolor{g1}
\rowcolor{g1}\bf{Center}&\bf{Dice(\%)} & \bf{Jaccard(\%)} & \bf{Precision(\%)} &\bf{Recall(\%)} & \bf{HD95(mm)} & \bf{ASSD(mm)} \\\hline
\rowcolor{g3}\multicolumn{7}{l}{T1W Modality}\\\hline
\rowcolor{g2}NYU&70.36$\pm$2.53 & 56.56$\pm$2.52 & 69.05$\pm$3.13 & 75.88$\pm$1.13 & 45.50$\pm$9.37 & 14.81$\pm$5.80\\
\rowcolor{g2}MCF&78.06$\pm$3.01 & 65.45$\pm$3.43 & 78.71$\pm$2.11 & 79.71$\pm$3.99 & 16.84$\pm$4.69 & 3.89$\pm$1.76\\
\rowcolor{g2}NU&77.12$\pm$3.06 & 64.79$\pm$3.77 & 73.90$\pm$3.36 & 83.19$\pm$2.12 & 34.39$\pm$8.18 & 8.86$\pm$2.00\\
\rowcolor{g2}AHN&49.40$\pm$10.96 & 36.07$\pm$8.83 & 58.90$\pm$8.21 & 46.13$\pm$12.58 & 44.40$\pm$25.11 & 14.65$\pm$9.00\\
\rowcolor{g2}MCA&41.21$\pm$13.62 & 28.85$\pm$11.78 & 37.23$\pm$12.92 & 54.25$\pm$14.72 & 118.45$\pm$41.69 & 53.98$\pm$23.47\\
\rowcolor{g2}IU&68.46$\pm$3.17 & 55.10$\pm$3.98 & 64.54$\pm$3.92 & 77.89$\pm$4.16 & 62.72$\pm$7.43 & 19.55$\pm$3.45\\
\bottomrule
    \end{tabular}
    \label{tab: SwinUnetRFedProx0005}
\end{table}

\begin{table}[htbp]
\caption{{T1W IPMN-Only Binary Classification.}}
\centering
\begin{tiny}
\aboverulesep=0ex
\belowrulesep=0ex
\setlength{\tabcolsep}{1pt}
{
\begin{tabular}{l|ll|lll|lll}
\toprule
\belowrulesepcolor{g1}
\rowcolor{g1}&&&\multicolumn{3}{l|}{\textbf{Validation-Selected Threshold}}&\multicolumn{3}{l}{\textbf{Fixed 0.5 Threshold}}\\
\rowcolor{g1}\multirow{-2}{*}{\textbf{Method}}&\multirow{-2}{*}{\textbf{AUC(\%)}}&\multirow{-2}{*}{\textbf{95\%CI(\%)}}&\textbf{ACC(\%)}&\textbf{Sens(\%)}&\textbf{Spec(\%)}&\textbf{ACC(\%)}&\textbf{Sens(\%)}&\textbf{Spec(\%)}\\
\hline
\rowcolor{g3}\multicolumn{9}{l}{\textbf{Internal 5-fold test evaluation, 339 low-risk + 149 high-risk; results are mean $\pm$ s.d. across folds.}}\\
\hline
\rowcolor{g2}Radiomics & 73.19$\pm$2.55 & [68.21, 78.30] & 73.96$\pm$2.79 & 44.83$\pm$9.29 & 86.63$\pm$6.82 & 73.34$\pm$3.50 & 35.04$\pm$6.80 & 90.23$\pm$2.66 \\
\rowcolor{g2}DenseNet-121 & 74.59$\pm$3.28 & [69.38, 79.76] & 73.35$\pm$2.74 & 43.47$\pm$6.64 & 86.33$\pm$6.11 & 70.51$\pm$3.16 & 51.26$\pm$22.49 & 79.24$\pm$12.49 \\
\hline
\rowcolor{g1}\multicolumn{9}{l}{\textbf{External leave-one-center-out evaluation: Train on six centers and test on the remaining one}}\\
\hline
\rowcolor{g3}\multicolumn{9}{l}{\textbf{Center 1: New York University Langone Health (NYU), 79 low-risk + 23 high-risk}}\\
\hline
\rowcolor{g2}Radiomics & 74.46 & [64.00, 84.21] & 75.49 & 34.78 & 87.34 & 72.55 & 56.52 & 77.22 \\
\bottomrule
\end{tabular}}
\end{tiny}
\label{tab: t1 low vs high classification}
\end{table}

\begin{table}[htbp]
\caption{{T2W IPMN-Only Binary Classification.}}
\centering
\begin{tiny}
\aboverulesep=0ex
\belowrulesep=0ex
\setlength{\tabcolsep}{1pt}
{
\begin{tabular}{l|ll|lll|lll}
\toprule
\belowrulesepcolor{g1}
\rowcolor{g1}&&&\multicolumn{3}{l|}{\textbf{Validation-Selected Threshold}}&\multicolumn{3}{l}{\textbf{Fixed 0.5 Threshold}}\\
\rowcolor{g1}\multirow{-2}{*}{\textbf{Method}}&\multirow{-2}{*}{\textbf{AUC(\%)}}&\multirow{-2}{*}{\textbf{95\%CI(\%)}}&\textbf{ACC(\%)}&\textbf{Sens(\%)}&\textbf{Spec(\%)}&\textbf{ACC(\%)}&\textbf{Sens(\%)}&\textbf{Spec(\%)}\\
\hline
\rowcolor{g3}\multicolumn{9}{l}{\textbf{Internal 5-fold test evaluation, 344 low-risk + 152 high-risk; results are mean $\pm$ s.d. across folds.}}\\
\hline
\rowcolor{g2}Radiomics & 73.64$\pm$2.91 & [68.93, 78.18] & 73.19$\pm$2.38 & 40.09$\pm$11.99 & 87.84$\pm$6.39 & 74.59$\pm$2.78 & 31.57$\pm$3.85 & 93.62$\pm$3.16 \\
\rowcolor{g2}DenseNet-121 & 73.45$\pm$5.28 & [68.89, 78.03] & 73.39$\pm$3.91 & 44.71$\pm$10.02 & 86.06$\pm$3.87 & 71.01$\pm$4.63 & 43.29$\pm$31.08 & 83.29$\pm$15.70 \\
\hline
\rowcolor{g1}\multicolumn{9}{l}{\textbf{External leave-one-center-out evaluation: Train on six centers and test on the remaining one}}\\
\hline
\rowcolor{g3}\multicolumn{9}{l}{\textbf{Center 1: New York University Langone Health (NYU), 79 low-risk + 24 high-risk}}\\
\hline
\rowcolor{g2}Radiomics & 75.26 & [64.51, 84.82] & 79.61 & 25.00 & 96.20 & 78.64 & 25.00 & 94.94 \\
\bottomrule
\end{tabular}}
\end{tiny}
\label{tab: t2 low vs high classification}
\end{table}

\begin{table}[htbp]
\caption{{\textit{DenseNet-121} Two-Modality IPMN-Only Binary Classification.}}
\centering
\begin{tiny}
\aboverulesep=0ex
\belowrulesep=0ex
\setlength{\tabcolsep}{1pt}
{
\begin{tabular}{l|ll|lll|lll}
\toprule
\belowrulesepcolor{g1}
\rowcolor{g1}&&&\multicolumn{3}{l|}{\textbf{Validation-Selected Threshold}}&\multicolumn{3}{l}{\textbf{Fixed 0.5 Threshold}}\\
\rowcolor{g1}\multirow{-2}{*}{\textbf{Method}}&\multirow{-2}{*}{\textbf{AUC(\%)}}&\multirow{-2}{*}{\textbf{95\%CI(\%)}}&\textbf{ACC(\%)}&\textbf{Sens(\%)}&\textbf{Spec(\%)}&\textbf{ACC(\%)}&\textbf{Sens(\%)}&\textbf{Spec(\%)}\\
\hline
\rowcolor{g3}\multicolumn{9}{l}{\textbf{Internal 5-fold test evaluation, 332 low-risk + 142 high-risk; results are mean $\pm$ s.d. across folds.}}\\
\hline
\rowcolor{g2}Early fusion & 77.19$\pm$4.38 & [72.77, 81.36] & 74.50$\pm$2.98 & 45.17$\pm$4.91 & 87.11$\pm$4.00 & 75.16$\pm$3.99 & 43.02$\pm$20.19 & 88.86$\pm$5.11 \\
\rowcolor{g2}Late fusion & 73.25$\pm$5.54 & [68.52, 78.34] & 72.38$\pm$3.92 & 42.23$\pm$9.19 & 85.33$\pm$6.10 & 67.12$\pm$8.53 & 55.41$\pm$23.63 & 72.60$\pm$21.20 \\
\rowcolor{g2}Siamese fusion & 73.91$\pm$6.20 & [68.90, 79.07] & 71.50$\pm$4.10 & 40.04$\pm$6.57 & 84.97$\pm$7.95 & 70.47$\pm$3.22 & 39.15$\pm$24.69 & 84.14$\pm$14.82 \\
\rowcolor{g2}Logit fusion & 71.51$\pm$3.21 & [66.22, 76.67] & 71.08$\pm$1.67 & 40.59$\pm$11.40 & 84.03$\pm$5.56 & 72.97$\pm$1.50 & 18.14$\pm$9.58 & 96.45$\pm$2.98 \\
\hline
\rowcolor{g1}\multicolumn{9}{l}{\textbf{External leave-one-center-out evaluation: Train on six centers and test on the remaining one}}\\
\bottomrule
\end{tabular}}
\end{tiny}
\label{tab: fusion low vs high}
\end{table}

\begin{table}[htbp]
\caption{{T1W Broad Binary IPMN Classification.}}
\centering
\begin{tiny}
\aboverulesep=0ex
\belowrulesep=0ex
\setlength{\tabcolsep}{1pt}
{
\begin{tabular}{l|ll|lll|lll}
\toprule
\belowrulesepcolor{g1}
\rowcolor{g1}&&&\multicolumn{3}{l|}{\textbf{Validation-Selected Threshold}}&\multicolumn{3}{l}{\textbf{Fixed 0.5 Threshold}}\\
\rowcolor{g1}\multirow{-2}{*}{\textbf{Method}}&\multirow{-2}{*}{\textbf{AUC(\%)}} & \multirow{-2}{*}{\textbf{95\%CI(\%)}} & \textbf{ACC(\%)}&\textbf{Sens(\%)} & \textbf{Spec(\%)}& \textbf{ACC(\%)}&\textbf{Sens(\%)} & \textbf{Spec(\%)} \\
\hline
\rowcolor{g3}\multicolumn{9}{l}{\textbf{Internal 5-fold evaluation, 503 non-high-risk + 149 high-risk; results are mean $\pm$ s.d. across folds.}}\\
\hline
\rowcolor{g1}Radiomics & 78.02$\pm$0.97 & [73.45, 82.06] & 79.14$\pm$1.17 & 40.27$\pm$6.31 & 90.63$\pm$2.23 & 80.22$\pm$1.44 & 23.57$\pm$9.06 & 96.99$\pm$1.71 \\
\rowcolor{g2}ResNet-34 & 76.72$\pm$1.65 & [72.16, 80.89] & 77.90$\pm$1.85 & 42.76$\pm$11.05 & 88.27$\pm$2.70 & 75.03$\pm$3.38 & 35.79$\pm$24.54 & 86.77$\pm$10.76 \\
\rowcolor{g2}ResNet-50 & 77.83$\pm$2.50 & [73.48, 81.69] & 77.90$\pm$3.44 & 48.94$\pm$7.51 & 86.40$\pm$5.78 & 77.00$\pm$2.33 & 44.32$\pm$15.57 & 86.68$\pm$6.66 \\
\rowcolor{g2}EfficientNet-B0 & 76.33$\pm$5.34 & [71.69, 80.62] & 77.79$\pm$2.63 & 40.23$\pm$11.23 & 88.87$\pm$3.60 & 73.14$\pm$4.83 & 52.64$\pm$25.80 & 79.32$\pm$13.02 \\
\rowcolor{g1}DenseNet-121 & 78.67$\pm$4.02 & [74.41, 82.50] & 79.28$\pm$3.18 & 46.96$\pm$2.48 & 88.83$\pm$3.77 & 74.83$\pm$7.15 & 53.65$\pm$21.56 & 81.09$\pm$14.81 \\
\hline
\end{tabular}
}
\end{tiny}
\label{tab: t1 classification}
\end{table}

\begin{table}[htbp]
\caption{{T2W Broad Binary IPMN Classification.}}
\centering
\begin{tiny}
\aboverulesep=0ex
\belowrulesep=0ex
\setlength{\tabcolsep}{1pt}
{
\begin{tabular}{l|ll|lll|lll}
\toprule
\belowrulesepcolor{g1}
\rowcolor{g1}&&&\multicolumn{3}{l|}{\textbf{Validation-Selected Threshold}}&\multicolumn{3}{l}{\textbf{Fixed 0.5 Threshold}}\\
\rowcolor{g1}\multirow{-2}{*}{\textbf{Method}}&\multirow{-2}{*}{\textbf{AUC(\%)}} & \multirow{-2}{*}{\textbf{95\%CI(\%)}} & \textbf{ACC(\%)}&\textbf{Sens(\%)} & \textbf{Spec(\%)}& \textbf{ACC(\%)}&\textbf{Sens(\%)} & \textbf{Spec(\%)} \\
\hline
\rowcolor{g3}\multicolumn{9}{l}{\textbf{Internal 5-fold evaluation, 503 non-high-risk + 152 high-risk; results are mean $\pm$ s.d. across folds.}}\\
\hline
\rowcolor{g1}Radiomics & 78.91$\pm$1.71 & [74.20, 82.86] & 80.00$\pm$1.48 & 46.13$\pm$10.15 & 90.35$\pm$3.43 & 80.15$\pm$1.98 & 27.10$\pm$9.25 & 96.27$\pm$1.98 \\
\rowcolor{g2}ResNet-34 & 79.44$\pm$2.73 & [75.26, 83.54] & 79.08$\pm$0.84 & 42.15$\pm$7.61 & 90.22$\pm$2.77 & 78.94$\pm$1.08 & 33.70$\pm$12.17 & 92.74$\pm$4.93 \\
\rowcolor{g2}ResNet-50 & 78.96$\pm$2.82 & [74.80, 82.91] & 78.17$\pm$2.36 & 42.84$\pm$8.99 & 88.90$\pm$3.83 & 78.50$\pm$1.97 & 40.15$\pm$15.14 & 90.15$\pm$6.43 \\
\rowcolor{g2}EfficientNet-B0 & 77.10$\pm$2.10 & [72.38, 81.06] & 77.27$\pm$1.42 & 36.92$\pm$10.35 & 89.47$\pm$3.13 & 72.28$\pm$11.96 & 37.94$\pm$26.81 & 82.53$\pm$22.87 \\
\rowcolor{g1}DenseNet-121 & 81.28$\pm$2.44 & [77.28, 85.22] & 81.07$\pm$3.14 & 57.98$\pm$5.34 & 88.09$\pm$4.39 & 78.49$\pm$1.77 & 36.24$\pm$20.49 & 91.38$\pm$7.86 \\
\hline
\end{tabular}
}
\end{tiny}
\label{tab: t2 classification}
\end{table}

\begin{table}[htbp]
\caption{{{\textit{DenseNet-121} Two-Modality Broad Binary IPMN Classification.}}}
\centering
\begin{tiny}
\aboverulesep=0ex
\belowrulesep=0ex
\setlength{\tabcolsep}{1pt}
{
\begin{tabular}{l|ll|lll|lll}
\toprule
\belowrulesepcolor{g1}
\rowcolor{g1}&&&\multicolumn{3}{l|}{\textbf{Validation-Selected Threshold}}&\multicolumn{3}{l}{\textbf{Fixed 0.5 Threshold}}\\
\rowcolor{g1}\multirow{-2}{*}{\textbf{Method}}&\multirow{-2}{*}{\textbf{AUC(\%)}}  & \multirow{-2}{*}{\textbf{95\%CI(\%)}}& \textbf{ACC(\%)}&\textbf{Sens(\%)}& \textbf{Spec(\%)} & \textbf{ACC(\%)}&\textbf{Sens(\%)}& \textbf{Spec(\%)} \\
\hline
\rowcolor{g3}\multicolumn{9}{l}{\textbf{Internal 5-fold evaluation, 487 non-high-risk + 142 high-risk; results are mean $\pm$ s.d. across folds.}}\\
\hline
\rowcolor{g2}Early fusion & $80.79\pm5.44$ & [76.39, 84.84] & $80.62\pm3.20$ & $45.69\pm6.76$ & $90.78\pm2.78$ & $74.43\pm10.74$ & $65.10\pm17.54$ & $77.37\pm17.06$ \\
\rowcolor{g2}Late fusion & $78.06\pm6.57$ & [73.49, 81.99] & $77.69\pm2.78$ & $36.39\pm10.71$ & $89.74\pm2.19$ & $79.15\pm1.92$ & $16.38\pm14.35$ & $97.58\pm2.95$ \\
\rowcolor{g2}Siamese fusion & $81.24\pm4.78$ & [76.76, 84.99] & $79.37\pm2.51$ & $46.98\pm11.39$ & $88.72\pm3.92$ & $80.47\pm2.98$ & $47.74\pm22.47$ & $90.08\pm5.32$ \\
\rowcolor{g2}Logit fusion & $79.79\pm3.91$ & [75.63, 83.82] & $79.20\pm2.74$ & $45.02\pm7.96$ & $89.16\pm5.45$ & $74.42\pm10.81$ & $54.98\pm20.07$ & $80.29\pm18.58$ \\
\hline
\rowcolor{g1}\multicolumn{9}{l}{\textbf{External leave-one-center-out evaluation: Train on six centers and test on the remaining one}}\\
\bottomrule
\end{tabular}}
\end{tiny}
\label{tab: fusion full}
\end{table}

\begin{table}[p]
\centering
\caption{{Internal comparisons of task-specific T2W DenseNet-121 models with radiologists. Cohorts are the broad paired cohort ($n=629$), IPMN-only cohort ($n=474$), and histopathology-confirmed IPMN subset ($n=357$). Models use validation-selected thresholds. Values are percentages with conditional 95\% bootstrap confidence intervals.}}
\label{tab:zw-reader-ci}
{\small
\begin{tabular}{llll}
\toprule
Cohort & Method & Sensitivity [95\% CI] & Specificity [95\% CI]\\
\midrule
629 & T2W DenseNet & 57.04 [49.30, 64.79] & 88.50 [85.63, 91.38]\\
629 & Radiologist mean & 46.01 [39.20, 52.58] & 93.91 [92.20, 95.62]\\
474 & T2W DenseNet & 41.55 [33.80, 49.30] & 86.75 [82.83, 90.36]\\
474 & Radiologist mean & 46.01 [39.20, 52.58] & 91.67 [89.25, 93.78]\\
357 & T2W DenseNet & 30.28 [23.24, 38.05] & 89.77 [85.58, 93.49]\\
357 & Radiologist mean & 46.01 [39.20, 52.58] & 87.29 [83.41, 90.70]\\
\bottomrule
\end{tabular}}
\end{table}

\begin{table}[p]
\centering
\caption{{Paired differences between internal T2W DenseNet-121 and the mean across radiologists, in percentage points. Cohorts and operating thresholds are defined in Table~\ref{tab:zw-reader-ci}. Confidence intervals use the same 1,000 paired patient bootstrap replicates.}}
\label{tab:zw-paired-differences}
{\small
\begin{tabular}{llll}
\toprule
Cohort & Metric & AI minus radiologist mean & 95\% CI\\
\midrule
629 & Sensitivity & 11.03 & [3.05, 18.78]\\
629 & Specificity & $-$5.41 & [$-$8.08, $-$2.80]\\
474 & Sensitivity & $-$4.46 & [$-$13.15, 3.76]\\
474 & Specificity & $-$4.92 & [$-$8.43, $-$1.41]\\
357 & Sensitivity & $-$15.73 & [$-$23.94, $-$7.51]\\
357 & Specificity & 2.48 & [$-$2.48, 7.45]\\
\bottomrule
\end{tabular}}
\end{table}

\begin{table}[p]
\centering
\caption{{Evaluation of single-modality DenseNet-121 models on the same 357 histopathology-confirmed IPMN patients (142 high-risk, 215 low-risk). Broad and IPMN-only predictions and thresholds are frozen from the original tasks; histopathology-only rows use separately retrained models. Sens and Spec denote sensitivity and specificity at the original validation-selected thresholds. Values are percentages. AUCs use pooled held-out probabilities; their intervals are conditional 95\% bootstrap confidence intervals.}}
\label{tab:zw-frozen-pathology}
{\small
\begin{tabular}{llllll}
\toprule
Evaluation & Training task & MRI & Sens & Spec & AUC [95\% CI]\\
\midrule
Internal & Broad & T1W & 48.59 & 80.00 & 67.13 [61.39, 73.19]\\
Internal & Broad & T2W & 57.04 & 78.60 & 67.26 [61.58, 73.02]\\
Internal & IPMN-only & T1W & 45.77 & 81.40 & 66.51 [60.77, 72.61]\\
Internal & IPMN-only & T2W & 41.55 & 80.93 & 62.15 [56.23, 68.60]\\
Internal & Histopathology-only & T1W & 26.76 & 86.51 & 61.11 [55.05, 67.56]\\
Internal & Histopathology-only & T2W & 30.28 & 89.77 & 62.35 [56.23, 68.90]\\
External & Broad & T1W & 48.59 & 78.14 & 68.22 [62.45, 73.89]\\
External & Broad & T2W & 49.30 & 76.74 & 72.01 [66.70, 77.25]\\
External & IPMN-only & T1W & 36.62 & 84.19 & 62.21 [56.49, 68.31]\\
External & IPMN-only & T2W & 47.89 & 84.19 & 64.63 [58.98, 70.46]\\
External & Histopathology-only & T1W & 23.94 & 90.23 & 54.77 [48.29, 61.66]\\
External & Histopathology-only & T2W & 64.08 & 71.16 & 66.58 [60.96, 72.34]\\
\bottomrule
\end{tabular}}
\end{table}

\begin{table}[p]
\centering
\caption{{Matched DenseNet-121 AUC summaries for the broad task. Internal and external estimates use the same patients within each modality (652 T1W; 655 T2W). Center means are unweighted averages over the same seven centers. Int. and ext. denote internal cross-validation and external leave-one-center-out evaluation, respectively; FedProx values in parentheses specify $\mu$. Values are percentages. $\Delta$ denotes the difference in external AUC between each federated configuration and centralized training (federated minus centralized), in percentage points, with 95\% CIs from 1,000 bootstrap replicates in which patients were resampled within each center and true class (Section~\ref{sec:statistical analysis}).}}
\label{tab:zw-matched-auc}
{\footnotesize\setlength{\tabcolsep}{3pt}
\begin{tabular}{llllllll}
\toprule
Method & MRI & \shortstack[l]{Pooled\\int.} & \shortstack[l]{Pooled\\ext.} & \shortstack[l]{Center\\int.} & \shortstack[l]{Center\\ext.} & \shortstack[l]{$\Delta$ pooled ext.\\{[}95\% CI{]}} & \shortstack[l]{$\Delta$ center ext.\\{[}95\% CI{]}}\\
\midrule
Centralized & T1W & 74.94 & 71.35 & 68.74 & 74.22 & -- & --\\
Centralized & T2W & 78.17 & 77.49 & 68.57 & 71.18 & -- & --\\
FedAvg & T1W & 71.35 & 66.90 & 67.82 & 73.12 & $-$4.45 [$-$7.77, $-$1.04] & $-$1.10 [$-$7.22, 4.85]\\
FedAvg & T2W & 68.63 & 65.15 & 61.99 & 72.28 & $-$12.34 [$-$16.15, $-$8.64] & 1.10 [$-$4.54, 6.67]\\
FedProx (0.1) & T1W & 73.27 & 62.54 & 69.61 & 73.80 & $-$8.81 [$-$11.96, $-$5.74] & $-$0.42 [$-$4.32, 3.19]\\
FedProx (0.1) & T2W & 66.85 & 74.18 & 61.25 & 68.61 & $-$3.31 [$-$6.54, $-$0.28] & $-$2.57 [$-$8.15, 2.78]\\
FedProx (0.3) & T1W & 74.20 & 65.89 & 72.83 & 72.09 & $-$5.46 [$-$8.75, $-$2.27] & $-$2.13 [$-$6.90, 2.11]\\
FedProx (0.3) & T2W & 65.72 & 62.35 & 59.42 & 71.00 & $-$15.14 [$-$18.20, $-$12.09] & $-$0.18 [$-$5.26, 4.57]\\
\bottomrule
\end{tabular}}
\end{table}

\begin{table}[p]
\centering
\caption{{Calibration of pooled test predictions before and after validation-based recalibration. Each fold-specific (internal) or held-out-center-specific (external) model was recalibrated by logistic regression on the logit of its own validation predictions; test labels were not used. Slope and intercept denote the calibration slope (ideal 1) and calibration-in-the-large intercept (ideal 0) with 95\% CIs from 1,000 patient-level bootstrap replicates stratified by true class. Brier scores are shown before / after recalibration, with the prevalence-only reference in parentheses. Pooled AUCs are shown before / after recalibration. Int. and ext. denote internal cross-validation and external leave-one-center-out evaluation; FedProx values in parentheses specify \mbox{$\mu$}.}}
\label{tab:zw-calibration}
{\tiny
\setlength{\tabcolsep}{2.5pt}
\begin{tabular}{llllllllll}
\toprule
 & & & & \multicolumn{2}{c}{Before recalibration} & \multicolumn{2}{c}{After recalibration} & & \\
\cmidrule(lr){5-6}\cmidrule(lr){7-8}
Setting & Task & Model & MRI & Slope & Intercept & Slope & Intercept & Brier & Pooled AUC (\%)\\
\midrule
Int. & IPMN-only & DenseNet-121 & T1W & 0.42 [0.30, 0.56] & $-$0.28 [$-$0.44, $-$0.11] & 0.59 [0.46, 0.74] & $-$0.19 [$-$0.34, $-$0.05] & 0.203 / 0.189 (0.212) & 70.84 / 73.74\\
Int. & IPMN-only & DenseNet-121 & T2W & 0.19 [0.12, 0.27] & 0.47 [0.16, 0.76] & 0.74 [0.56, 0.94] & $-$0.19 [$-$0.29, $-$0.08] & 0.232 / 0.186 (0.213) & 67.86 / 73.61\\
Int. & IPMN-only & Radiomics & T1W & 0.95 [0.73, 1.22] & $-$0.02 [$-$0.10, 0.07] & 0.79 [0.60, 1.02] & $-$0.10 [$-$0.20, 0.00] & 0.181 / 0.183 (0.212) & 72.96 / 73.10\\
Int. & IPMN-only & Radiomics & T2W & 0.99 [0.72, 1.28] & 0.03 [$-$0.06, 0.10] & 0.58 [0.42, 0.75] & $-$0.22 [$-$0.37, $-$0.10] & 0.180 / 0.189 (0.213) & 73.55 / 72.90\\
Int. & Broad & DenseNet-121 & T1W & 0.39 [0.30, 0.48] & $-$0.55 [$-$0.76, $-$0.33] & 0.69 [0.57, 0.85] & $-$0.06 [$-$0.20, 0.08] & 0.177 / 0.148 (0.176) & 74.94 / 78.46\\
Int. & Broad & DenseNet-121 & T2W & 0.52 [0.42, 0.64] & 0.16 [$-$0.02, 0.33] & 0.76 [0.63, 0.95] & $-$0.03 [$-$0.17, 0.10] & 0.155 / 0.134 (0.178) & 78.17 / 80.92\\
Int. & Broad & FedAvg & T1W & 0.35 [0.27, 0.45] & 1.11 [0.92, 1.33] & 0.85 [0.70, 1.05] & 0.03 [$-$0.08, 0.14] & 0.166 / 0.143 (0.176) & 71.35 / 77.67\\
Int. & Broad & FedAvg & T2W & 0.29 [0.20, 0.39] & 0.38 [0.25, 0.50] & 0.76 [0.60, 0.93] & $-$0.11 [$-$0.21, 0.00] & 0.171 / 0.153 (0.178) & 68.63 / 75.74\\
Int. & Broad & FedProx (0.1) & T1W & 0.35 [0.26, 0.45] & 0.25 [0.07, 0.41] & 0.75 [0.60, 0.93] & 0.04 [$-$0.08, 0.16] & 0.162 / 0.142 (0.176) & 73.27 / 78.56\\
Int. & Broad & FedProx (0.1) & T2W & 0.40 [0.28, 0.53] & 0.05 [$-$0.07, 0.17] & 0.75 [0.60, 0.92] & $-$0.01 [$-$0.14, 0.10] & 0.182 / 0.149 (0.178) & 66.85 / 77.27\\
Int. & Broad & FedProx (0.3) & T1W & 0.69 [0.52, 0.88] & 0.22 [0.11, 0.32] & 0.71 [0.56, 0.88] & 0.01 [$-$0.11, 0.13] & 0.151 / 0.145 (0.176) & 74.20 / 77.58\\
Int. & Broad & FedProx (0.3) & T2W & 0.14 [0.10, 0.20] & 0.42 [0.17, 0.65] & 0.78 [0.63, 0.97] & $-$0.09 [$-$0.21, 0.03] & 0.198 / 0.146 (0.178) & 65.72 / 79.02\\
Ext. & IPMN-only & DenseNet-121 & T1W & 0.24 [0.17, 0.32] & $-$0.34 [$-$0.61, $-$0.08] & 0.43 [0.28, 0.61] & $-$0.11 [$-$0.23, 0.01] & 0.251 / 0.215 (0.212) & 66.93 / 64.44\\
Ext. & IPMN-only & DenseNet-121 & T2W & 0.30 [0.21, 0.41] & $-$0.66 [$-$0.85, $-$0.47] & 0.58 [0.43, 0.75] & $-$0.07 [$-$0.22, 0.07] & 0.232 / 0.189 (0.213) & 69.30 / 72.61\\
Ext. & IPMN-only & Radiomics & T1W & 0.61 [0.36, 0.89] & $-$0.08 [$-$0.14, 0.00] & 0.53 [0.28, 0.79] & $-$0.07 [$-$0.14, 0.00] & 0.204 / 0.206 (0.212) & 64.56 / 63.52\\
Ext. & IPMN-only & Radiomics & T2W & 0.81 [0.52, 1.14] & 0.00 [$-$0.06, 0.06] & 0.66 [0.45, 0.89] & 0.26 [0.16, 0.35] & 0.199 / 0.196 (0.213) & 64.66 / 69.28\\
Ext. & Broad & DenseNet-121 & T1W & 0.30 [0.17, 0.45] & $-$0.78 [$-$0.94, $-$0.62] & 0.70 [0.56, 0.89] & $-$0.06 [$-$0.19, 0.06] & 0.190 / 0.153 (0.176) & 71.35 / 76.99\\
Ext. & Broad & DenseNet-121 & T2W & 0.40 [0.31, 0.50] & $-$0.72 [$-$0.90, $-$0.52] & 0.74 [0.56, 0.94] & 0.00 [$-$0.10, 0.10] & 0.175 / 0.154 (0.178) & 77.49 / 75.56\\
Ext. & Broad & FedAvg & T1W & 0.34 [0.23, 0.47] & $-$0.37 [$-$0.49, $-$0.25] & 0.60 [0.45, 0.76] & $-$0.15 [$-$0.25, $-$0.04] & 0.195 / 0.165 (0.176) & 66.90 / 71.67\\
Ext. & Broad & FedAvg & T2W & 0.22 [0.13, 0.34] & 0.00 [$-$0.14, 0.14] & 0.60 [0.46, 0.76] & 0.10 [$-$0.02, 0.22] & 0.208 / 0.162 (0.178) & 65.15 / 72.05\\
Ext. & Broad & FedProx (0.1) & T1W & 0.24 [0.13, 0.36] & $-$0.26 [$-$0.41, $-$0.13] & 0.62 [0.47, 0.79] & $-$0.07 [$-$0.19, 0.04] & 0.203 / 0.160 (0.176) & 62.54 / 73.99\\
Ext. & Broad & FedProx (0.1) & T2W & 0.49 [0.39, 0.60] & $-$0.23 [$-$0.38, $-$0.08] & 0.78 [0.62, 0.95] & 0.02 [$-$0.08, 0.12] & 0.168 / 0.153 (0.178) & 74.18 / 75.38\\
Ext. & Broad & FedProx (0.3) & T1W & 0.30 [0.20, 0.43] & $-$1.25 [$-$1.37, $-$1.13] & 0.58 [0.44, 0.74] & $-$0.09 [$-$0.20, 0.03] & 0.238 / 0.158 (0.176) & 65.89 / 74.11\\
Ext. & Broad & FedProx (0.3) & T2W & 0.22 [0.16, 0.29] & $-$0.36 [$-$0.49, $-$0.22] & 0.81 [0.66, 1.01] & 0.07 [$-$0.03, 0.18] & 0.215 / 0.147 (0.178) & 62.35 / 78.17\\
\bottomrule
\end{tabular}}
\end{table}

\begin{table}[p]
\centering
\caption{{Decision-curve analysis of the recalibrated models. Net benefit at threshold probabilities of 0.10--0.40 is shown, with the net benefit of treating all patients in parentheses; the net benefit of treating no patients is zero. \mbox{$^{*}$}The 95\% CI of the difference between the model and the better of the two reference strategies excluded zero (1,000 patient-level bootstrap replicates stratified by true class). Int. and ext. denote internal cross-validation and external leave-one-center-out evaluation; FedProx values in parentheses specify \mbox{$\mu$}.}}
\label{tab:zw-netbenefit}
{\tiny
\setlength{\tabcolsep}{3pt}
\begin{tabular}{llllllll}
\toprule
Setting & Task & Model & MRI & 0.10 & 0.20 & 0.30 & 0.40\\
\midrule
Int. & IPMN-only & DenseNet-121 & T1W & 0.230 (0.228) & 0.162$^{*}$ (0.132) & 0.114$^{*}$ (0.008) & 0.072$^{*}$ ($-$0.158)\\
Int. & IPMN-only & DenseNet-121 & T2W & 0.232 (0.229) & 0.171$^{*}$ (0.133) & 0.108$^{*}$ (0.009) & 0.056$^{*}$ ($-$0.156)\\
Int. & IPMN-only & Radiomics & T1W & 0.229 (0.228) & 0.161$^{*}$ (0.132) & 0.097$^{*}$ (0.008) & 0.067$^{*}$ ($-$0.158)\\
Int. & IPMN-only & Radiomics & T2W & 0.229 (0.229) & 0.160$^{*}$ (0.133) & 0.099$^{*}$ (0.009) & 0.058$^{*}$ ($-$0.156)\\
Int. & Broad & DenseNet-121 & T1W & 0.155 (0.143) & 0.110$^{*}$ (0.036) & 0.074$^{*}$ ($-$0.102) & 0.047$^{*}$ ($-$0.286)\\
Int. & Broad & DenseNet-121 & T2W & 0.160 (0.147) & 0.117$^{*}$ (0.040) & 0.085$^{*}$ ($-$0.097) & 0.069$^{*}$ ($-$0.280)\\
Int. & Broad & FedAvg & T1W & 0.156$^{*}$ (0.143) & 0.098$^{*}$ (0.036) & 0.073$^{*}$ ($-$0.102) & 0.042$^{*}$ ($-$0.286)\\
Int. & Broad & FedAvg & T2W & 0.154 (0.147) & 0.103$^{*}$ (0.040) & 0.066$^{*}$ ($-$0.097) & 0.030$^{*}$ ($-$0.280)\\
Int. & Broad & FedProx (0.1) & T1W & 0.152 (0.143) & 0.109$^{*}$ (0.036) & 0.080$^{*}$ ($-$0.102) & 0.052$^{*}$ ($-$0.286)\\
Int. & Broad & FedProx (0.1) & T2W & 0.159 (0.147) & 0.104$^{*}$ (0.040) & 0.074$^{*}$ ($-$0.097) & 0.043$^{*}$ ($-$0.280)\\
Int. & Broad & FedProx (0.3) & T1W & 0.153 (0.143) & 0.107$^{*}$ (0.036) & 0.076$^{*}$ ($-$0.102) & 0.044$^{*}$ ($-$0.286)\\
Int. & Broad & FedProx (0.3) & T2W & 0.167$^{*}$ (0.147) & 0.118$^{*}$ (0.040) & 0.085$^{*}$ ($-$0.097) & 0.044$^{*}$ ($-$0.280)\\
Ext. & IPMN-only & DenseNet-121 & T1W & 0.227 (0.228) & 0.132 (0.132) & 0.073$^{*}$ (0.008) & 0.004 ($-$0.158)\\
Ext. & IPMN-only & DenseNet-121 & T2W & 0.224 (0.229) & 0.151 (0.133) & 0.097$^{*}$ (0.009) & 0.065$^{*}$ ($-$0.156)\\
Ext. & IPMN-only & Radiomics & T1W & 0.215 (0.228) & 0.124 (0.132) & 0.061$^{*}$ (0.008) & 0.036$^{*}$ ($-$0.158)\\
Ext. & IPMN-only & Radiomics & T2W & 0.226 (0.229) & 0.138 (0.133) & 0.097$^{*}$ (0.009) & 0.065$^{*}$ ($-$0.156)\\
Ext. & Broad & DenseNet-121 & T1W & 0.157$^{*}$ (0.143) & 0.111$^{*}$ (0.036) & 0.063$^{*}$ ($-$0.102) & 0.034$^{*}$ ($-$0.286)\\
Ext. & Broad & DenseNet-121 & T2W & 0.163$^{*}$ (0.147) & 0.099$^{*}$ (0.040) & 0.070$^{*}$ ($-$0.097) & 0.025$^{*}$ ($-$0.280)\\
Ext. & Broad & FedAvg & T1W & 0.149 (0.143) & 0.079$^{*}$ (0.036) & 0.043$^{*}$ ($-$0.102) & 0.025$^{*}$ ($-$0.286)\\
Ext. & Broad & FedAvg & T2W & 0.150 (0.147) & 0.073$^{*}$ (0.040) & 0.047$^{*}$ ($-$0.097) & 0.031$^{*}$ ($-$0.280)\\
Ext. & Broad & FedProx (0.1) & T1W & 0.156$^{*}$ (0.143) & 0.099$^{*}$ (0.036) & 0.045$^{*}$ ($-$0.102) & 0.026$^{*}$ ($-$0.286)\\
Ext. & Broad & FedProx (0.1) & T2W & 0.162$^{*}$ (0.147) & 0.097$^{*}$ (0.040) & 0.055$^{*}$ ($-$0.097) & 0.039$^{*}$ ($-$0.280)\\
Ext. & Broad & FedProx (0.3) & T1W & 0.150 (0.143) & 0.092$^{*}$ (0.036) & 0.053$^{*}$ ($-$0.102) & 0.029$^{*}$ ($-$0.286)\\
Ext. & Broad & FedProx (0.3) & T2W & 0.157 (0.147) & 0.119$^{*}$ (0.040) & 0.074$^{*}$ ($-$0.097) & 0.043$^{*}$ ($-$0.280)\\
\bottomrule
\end{tabular}}
\end{table}

\clearpage




\end{appendices}

\end{document}